\documentclass[aps,prb,twocolumn,shortbibliography,superscriptaddress,article]{revtex4-1}
\usepackage{epsfig}
\usepackage{epstopdf}
\usepackage{amsmath}
\usepackage{amsfonts}
\usepackage{amssymb}
\usepackage{hyperref}
\usepackage{bm}
\usepackage{makecell}
\usepackage{rotating}
\usepackage{hyperref}
\usepackage{multirow}
\usepackage{graphicx}
\usepackage{array}
\usepackage{placeins}
\usepackage{tabularx} % make sure this is in your preamble
\newcolumntype{Y}{>{\centering\arraybackslash}X}

\usepackage{graphicx}% Include figure files
\usepackage{dcolumn}% Align table columns on the decimal point
\usepackage{bm}% bold math
\usepackage{color}

\usepackage{tikz,xcolor,hyperref}

\definecolor{lime}{HTML}{A6CE39}
\DeclareRobustCommand{\orcidicon}{%
	\begin{tikzpicture}
	\draw[lime, fill=lime] (0,0)
	circle [radius=0.16]
	node[white] {{\fontfamily{qag}\selectfont \tiny ID}};
	\draw[white, fill=white] (-0.0625,0.095)
	circle [radius=0.007];
	\end{tikzpicture}
	\hspace{-2mm}
}

\foreach \x in {A, ..., Z}{%
	\expandafter\xdef\csname orcid\x\endcsname{\noexpand\href{https://orcid.org/\csname orcidauthor\x\endcsname}{\noexpand\orcidicon}}
}

\begin{document}

\title{Symmetry-protected nodal planes and accidental nodal surfaces \\ in mixed odd-even wave spin-momentum locking of relativistic altermagnets}

\author{Xujia Gong\orcidA}
\email{xgong@magtop.ifpan.edu.pl}
\affiliation{International Research Centre Magtop, Institute of Physics, Polish Academy of Sciences, Aleja Lotnik\'ow 32/46, PL-02668 Warsaw, Poland}

\author{Amar Fakhredine\orcidF}
\email{amarf@ifpan.edu.pl}
\affiliation{Institute of Physics, Polish Academy of Sciences, Aleja Lotnik\'ow 32/46, 02668 Warsaw, Poland}

\author{Sahar Izadi Vishkayi\orcidS}
\email{izadi.123@gmail.com}
\affiliation{School of Quantum Physics and Matter, Institute for Research in Fundamental Sciences (IPM), P. O. Box 19395-5531, Tehran, Iran}

\author{Carmine Autieri\orcidB}
\email{autieri@magtop.ifpan.edu.pl}
\affiliation{International Research Centre Magtop, Institute of Physics, Polish Academy of Sciences,
Aleja Lotnik\'ow 32/46, PL-02668 Warsaw, Poland}
\affiliation{SPIN-CNR, UOS Salerno, IT-84084 Fisciano (SA), Italy}

\date{\today}

%\section*{Significance}
%Altermagnets host unconventional spin–momentum locking characterized by symmetry-enforced nodal planes in momentum space. Here we show that relativistic effects, including spin–orbit coupling and inversion-symmetry breaking, qualitatively modify this structure by reducing symmetry and reshaping nodal manifolds. In particular, we identify conditions under which g-wave spin textures survive in the relativistic limit and demonstrate the emergence of p-wave magnetism with symmetry-protected nodal planes alongside accidental nodal surfaces. These results establish that relativistic altermagnets support multiple coexisting angular-momentum spin textures and distinct nodal topologies. Our work provides a unified framework for understanding spin–momentum locking beyond the non-relativistic limit and opens routes to symmetry-based design of spintronic functionalities.

\begin{abstract} % done
% introduction
Non-relativistic spin--momentum locking in altermagnets exhibits an even number of nodal planes. In the relativistic limit, the number of nodal planes can be lowered by symmetry reduction due to the N\'eel vector and spin--orbit coupling in noncentrosymmetric systems. Therefore, an analysis of the evolution of the nodal planes in relativistic altermagnets is required.
%setting of the problem
While $g$-wave spin--momentum locking is straightforward to realize in non-relativistic altermagnets, this $g$-wave does not necessarily survive in the relativistic case. 
%A prototypical example is MnTe in the NiAs structure, which exhibits $g$-wave in the non-relativistic limit, whereas it displays $d$-wave once relativistic effects are included. 
In this work, we investigate the relativistic spin--momentum locking of the centrosymmetric CrSb and the noncentrosymmetric wurtzite MnTe.
%first result
As a first result, we show that in both systems the dominant spin component retains its $g$-wave character in the relativistic regime only when the N\'eel vector is oriented along the $z$-axis, while the subdominant components exhibit $d$-wave symmetry in CrSb and $p$-wave symmetry in ferroelectric wurtzite MnTe. More generally, the $g$-wave character is preserved in the relativistic limit only when both the N\'eel vector and the electric field associated with inversion-symmetry breaking are oriented along the $z$-axis.
%second result
As a second result, we show that relativistic spin--momentum locking of ferroelectric altermagnets can exhibit $p$-wave magnetism with one symmetry-protected nodal plane and an accidental nodal surface not protected by symmetry, or can have two accidental nodal surfaces. With the N\'eel vector aligned along the $x$-axis, selected bands of ferroelectric altermagnet wurtzite MnTe exhibit $p$-wave magnetism.
%Conclusions
Our results establish that altermagnets can host distinct spin components that realize a mixture of angular-momentum wave symmetries in momentum space in the relativistic limit. 
\end{abstract}

\maketitle	

\section{Introduction}
% altermagnets
In altermagnets, the sites with opposite spin are connected by rotational symmetries (proper or improper and symmorphic or nonsymmorphic) but not connected by translation or inversion symmetries\cite{Mazin2021,Smejkal22beyond,doi:10.1126/sciadv.aaz8809,hayami2019momentum,hayami2020bottom,Smejkal22,yuan2023degeneracy,Samanta2025,Sun2025,Xu2025,wei2024crystal,Zhang2025,D3NR03681B,D3NR04798A,ssxp-gz9l,Cuono23orbital,D4NR04053H,https://doi.org/10.1002/adfm.202505145,Berritta2025,jr65-4273,ref1}. While breaking of time-reversal symmetry and weak ferromagnetism induced by the spin-orbit coupling were already known for several decades\cite{DZYALOSHINSKY1958241}, one of the most striking novelties of the field of altermagnetism was the non-relativistic spin-momentum locking with even waves for magnetic systems\cite{Smejkal22beyond,song2025unifiedsymmetryclassificationmagnetic}.
The non-relativistic spin-momentum locking refers to a phenomenon where the electron’s spin orientation becomes tied (locked) to its crystal momentum due to rotational crystal symmetry. This assures a symmetry-protected zero net magnetization in the non-relativistic limit.  
%multipoles
Another recent development is the study of multipoles in magnetic systems\cite{doi:10.7566/JPSJ.93.072001,hayami2020bottom,PhysRevB.111.115150,han2025deterministicneelvectorswitching}, where the non-relativistic spin–momentum locking of altermagnets corresponds to quadrupole and higher-order multipole structures in $k$-space. Beyond their symmetry classification, altermagnets exhibit distinctive nonlinear response phenomena, such as photocurrent generation, which can be employed to probe magnetic structures and magnetoelectric switching\cite{yang2025nonlinear}, as well as ultrafast optical and valley-selective responses\cite{gao2025ultrafast}.

% altermagnetism + DMI
The spin-orbit coupling preserves the time-reversal symmetry. Therefore, in a system with Kramers degeneracy, the spin-orbit cannot create any net magnetization. In the case of altermagnetic compounds with broken time-reversal symmetry, the spin-orbit could generate the so-called weak ferromagnetism and anomalous Hall effect\cite{PhysRevLett.130.036702,PhysRevB.111.184407}. 
The presence of the altermagnetic spin-splitting is therefore a necessary condition to obtain weak ferromagnetism. Depending on the point group, different classes of altermagnets can produce weak ferromagnetism depending on the N\'eel vector orientation\cite{839n-rckn}, where the N\'eel vector is the difference between the spin vectors on the two magnetic sites and describes the direction of the spin. The simplest and most intuitive form of antisymmetric exchange is the staggered Dzyaloshinskii-Moriya interaction\cite{DZYALOSHINSKY1958241}, which produces relativistic weak ferromagnetism in altermagnets always orthogonal to the N\'eel vector\cite{PhysRevB.111.054442}. In chiral altermagnets, the absence of inversion and mirror symmetries allows spin–orbit coupling to generate persistent spin textures and, in combination with altermagnetic order, leads to Néel-vector–dependent spin transport signatures\cite{Tenzin2025,Gauswami2026}.

\begin{figure*}
    \centering
    \includegraphics[width=1\linewidth]{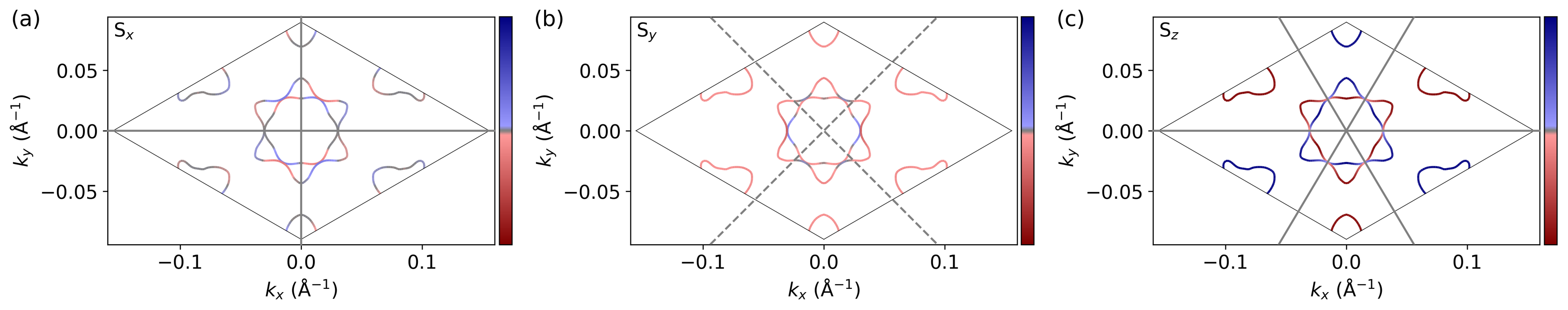}
    \caption{Center of the Brillouin zone in the $k_z$ = 0.25$\frac{\pi}{c}$ plane showing the two-dimensional Fermi surface of CrSb with N\'eel vector along the $z$-axis for (a) the $S_x$ component, (b) the $S_y$ component, and (c) the $S_z$ component. Red and blue denote bands with spectral weight of opposite spins.
    Combining our results at $k_z = 0$ and $k_z$ = 0.25$\frac{\pi}{c}$, the RSML of CrSb is composed of $Q_{xy}$, $Q_{xx}$-$Q_{yy}$ and $Q_{yz(3x^2-y^2)}$ for the $S_x$, $S_y$ and $S_z$ components, respectively. The solid black lines denote the nodal planes corresponding to each spin component; the dashed black line denotes the approximate position of the accidental nodal surfaces.}
    \label{fig:Figure1}
\end{figure*}

% RSML
The non-relativistic spin-momentum locking is inherited by the dominant spin component when spin-orbit coupling is included, whereas we refer to the other two as subdominant spin components. Beyond weak ferromagnetism, spin–orbit coupling has additional effects on the spin momentum locking of altermagnets, such as generating $d$-wave spin–momentum locking also for the subdominant components. As an extension of the concept of spin–momentum locking, we introduce the term relativistic spin–momentum locking (RSML) to denote the combination of spin–momentum locking for the three spin components.\cite{AutieriRSML,PhysRevB.109.024404,PhysRevB.110.144412}
% RSML beyond altermagnets
Beyond altermagnets, the RSML is also present in other systems breaking time-reversal symmetry, for instance, in the non-collinear MnTe$_2$\cite{Zhu2024} and in ferromagnets\cite{gong2026relativisticspinmomentumlockingferromagnets}.

%broken inversion symmetry without altermagnetism
When inversion symmetry is broken, spin–orbit interactions such as Rashba, Dresselhaus, or Weyl-type effects can become symmetry-allowed in materials\cite{yang2026symmetriesspinsplittinginducedspinorbit}. In particular, structural distortions either at interfaces or within the bulk can selectively generate these spin–orbit interactions, thereby controlling the resulting spin–momentum locking. In perovskite oxide heterostructures, recent studies have demonstrated that the specific form of spin–orbit coupling—Rashba, Dresselhaus, or their combination—can be directly linked to the nature of the underlying inversion asymmetry \cite{ganguli2025mapping}. 
% why RSML is important 
Materials with spin-momentum locking exhibit quantum metrics\cite{doi:10.1126/science.adq3255}. In systems where inversion symmetry is broken, symmetry-allowed Rashba or Dresselhaus spin-orbit interactions provide an additional route to engineer spin-momentum–locked states by structurally controlling the underlying asymmetry\cite{kawano2019designing}. When Rashba and Dresselhaus couplings coexist with comparable strength, they can stabilize a persistent spin helix state. The influence of the persistent spin-helix on the transport properties of altermagnets has also been investigated \cite{Tenzin2025,tm58-lbdl}. Another direction of research for the interplay between altermagnetism and breaking of inversion symmetry is the field of the ferroelectric altermagnets\cite{smejkal2024altermagneticmultiferroicsaltermagnetoelectriceffect,D4MH01619J,Bezzerga2025}.

In recent years, there has been growing interest in non-relativistic $p$-wave magnets with broken time-reversal symmetry. Achieving a pure $p$-wave magnetic state requires non-collinear magnetism, and several families of such magnets have been proposed in coplanar magnetic systems \cite{Chakraborty2025,mitscherling2026microscopicoriginpwavemagnetism,PhysRevLett.134.196907,PhysRevLett.133.236703,Song2025,vgcs-bn8g,zk69-k6b2,zhou2025anisotropicresistivitypwavemagnet,Yamada2025}. In contrast, the Rashba spin texture exhibits $p$-wave spin–momentum locking of relativistic origin, but it preserves time-reversal symmetry. We propose that the interplay between Rashba spin–orbit coupling and altermagnetism can induce $p$-wave magnetism with broken time-reversal symmetry, accompanied by other even-parity components, even in collinear magnetic systems.

%%%%%%%%%%% first result
In this paper, we will focus on the Rashba effect, but our results can be easily generalized to other spin-orbital effects such as Dresselhaus or persistent spin-helix. In previous manuscripts, we have shown examples of RSML with the combination of $p$-wave and $d$-wave spin-momentum locking\cite{leon2025strainenhancedaltermagnetismca3ru2o7,Fakhredine25b}, but not with the combination of $p$-wave and $g$-wave, which is trickier because the $g$-wave does not survive easily in the relativistic limit. 
% first result
By combining Rashba coupling with $g$-wave altermagnetism, we determine the conditions under which a magnetic system can host both $p$-wave and $g$-wave magnetic order that breaks time-reversal symmetry.
% second result
Additionally, we investigate the evolution of nodal planes in relativistic altermagnets as a function of the orientation of the N\'eel vector and of inversion-symmetry breaking. Our analysis starts from $g$-wave altermagnets, which represent a higher-order symmetry compared to $d$-wave altermagnets and can reduce to lower symmetries. We focus on two hexagonal space groups, considering one centrosymmetric and one noncentrosymmetric case. Among centrosymmetric materials, we study relativistic spin--momentum locking in the well-known altermagnet CrSb, which exhibits nonrelativistic $g$-wave spin--momentum locking \cite{Reimers2024} and topological Weyl phase\cite{doi:10.1021/acs.nanolett.5c00482,Li2025}. In the noncentrosymmetric case, we focus on the wurtzite phase of MnTe, which hosts ferroelectric properties\cite{Moriwake2020} and can be synthesized on suitable substrates with the wurtzite structure \cite{doi:10.1021/acsami.5c14582}. Recently, it was established that the isostructural wurtzite MnSe exhibits a larger internal electric field\cite{5f5l-7c8f}.

\begin{figure*}
    \centering    \includegraphics[width=1\linewidth]{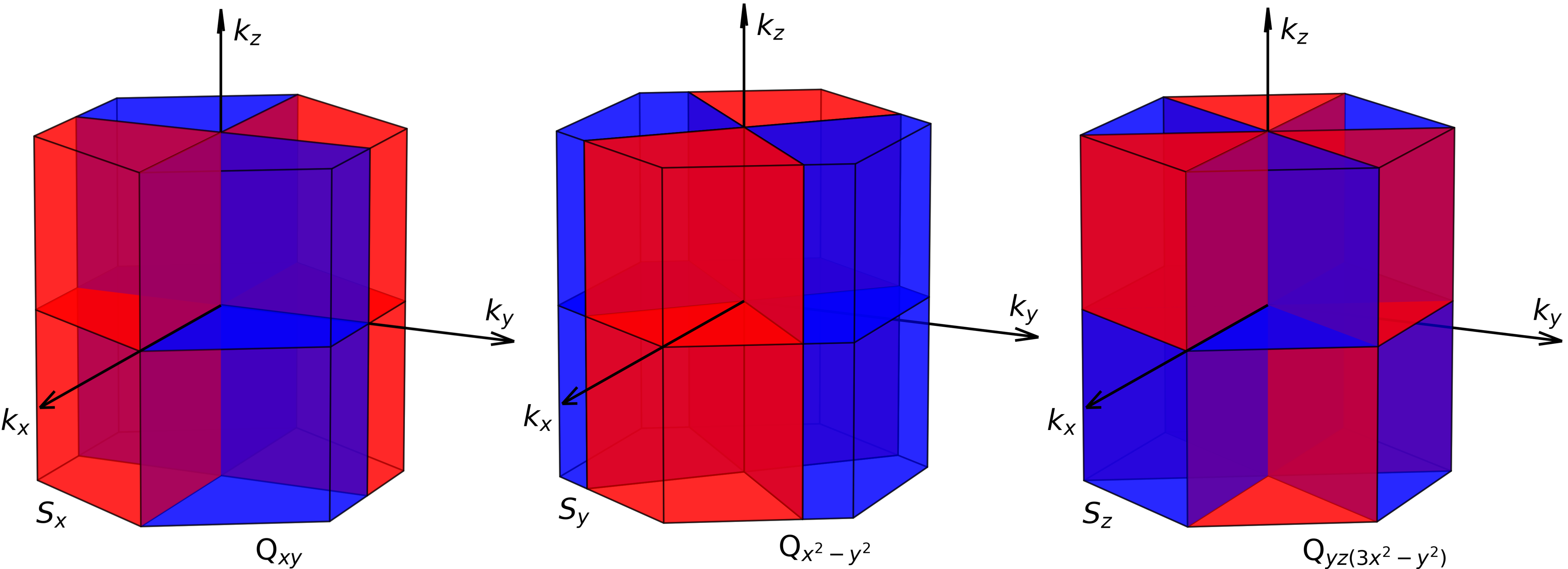}
    \caption{Relativistic spin-momentum locking of CrSb with N\'eel vector along the $z$-axis. The RSML is composed of $Q_{xy}$, $Q_{x^2-y^2}$ and $Q_{yz(3x^2-y^2)}$ for $S_x$, $S_y$ and $S_z$, respectively. Red and blue denote regions of the Brillouin zone with opposite spin-splitting. The SML of $S_x$ exhibits two nodal planes, the SML of $S_y$ exhibits 2 accidental nodal surfaces and the SML of $S_z$ exhibits 4 nodal planes.}
    \label{fig:Figure2}
\end{figure*} % Figure 2 - RSML of CrSb

% In this paper, we do.
This paper is organized as follows. In Sec.~II, we examine the relativistic spin--momentum locking in CrSb, analyze the resulting symmetry of the spin components in momentum space, and compare our results with previously reported calculations of relativistic spin--momentum locking in materials with the same space group but with the N\'eel vector oriented in the $xy$ plane. In Sec.~III, we perform a corresponding analysis for the altermagnetic phase of wurtzite MnTe with different N\'eel vectors. Therefore, in Section IV, we study the spin cantings of these material classes. Finally, in Sec.~V, we discuss the implications of our results, place them in the context of the existing literature, and present our conclusions.

\section{Relativistic spin-momentum locking of C\lowercase{r}S\lowercase{b}}

% general and MnTe
Compounds with the NiAs crystal structure belong to space group P6$_3$/mmc (no. 194). The properties of the crystal structure, together with the definitions of the axes in real and reciprocal space, are reported in the Supplementary Materials. Among them, two well-studied altermagnets are CrSb and MnTe\cite{kluczyk2023coexistence,Amin2024}.
The non-relativistic spin-momentum locking for both MnTe and CrSb is a $g$-wave with 4 nodal planes. The RSML depends on the N\'eel vector orientation. Experimentally, MnTe with NiAs structure exhibits a N\'eel vector along the $y$-axis\cite{PhysRevB.96.214418}, while CrSb exhibits a N\'eel vector along the $z$-axis.
The RSML for this class of materials is expressed in terms of magnetic quadrupoles for different N\'eel vector orientations and is reported in Table \ref{tab:CrSb_AM}, where the results for N\'eel vectors along the x- and y-axes are extracted from the literature  \cite{AutieriRSML,hirakida2025multipoleanalysisspincurrents}. The diagonal elements of the Table \ref{tab:CrSb_AM} report the dominant component. For the N\'eel vector along the x- or y-direction,  the spin-momentum locking is described by the magnetic quadrupole $Q_{yz}$ in $k$-space. For the magnetic quadrupole $Q_{yz}$, reversing the sign of the $k_y$ (or $k_z$) coordinate in the Brillouin zone produces an opposite spin-splitting for the $S_y$ component. Further details on the multipole notation and the procedure for extracting them from DFT calculations are provided in the Supplementary Materials. 

\begin{figure*}
    \centering
    \includegraphics[width=1\linewidth]{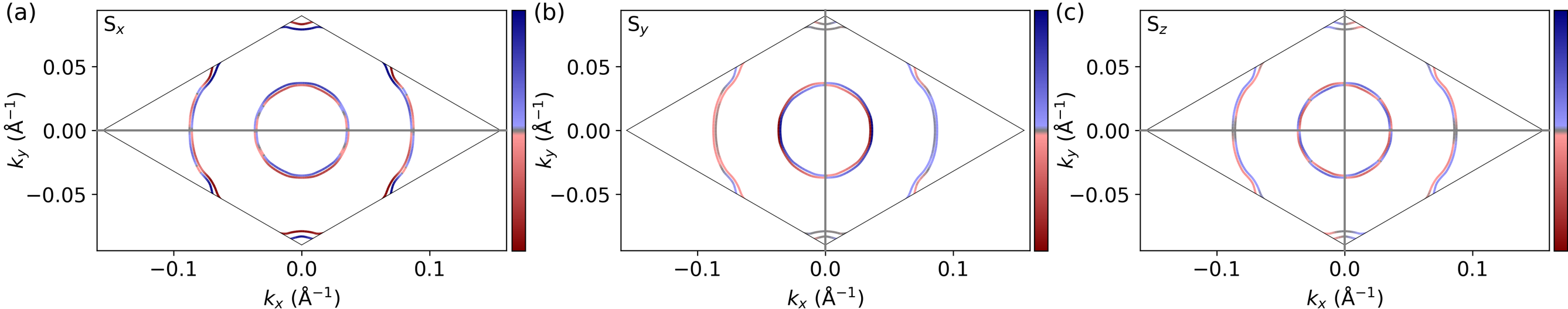}
    \caption{Center of the Brillouin zone in the $k_z = 0$ plane showing the two-dimensional Fermi surface 2.55 eV above the Fermi level of wurtzite MnTe with magnetization along the $x$-axis: (a) the $S_x$ component, (b) the $S_y$ component, and (c) the $S_z$ component. Red and blue denote bands with spectral weight of opposite spins. The solid black lines denote the symmetry-protected nodal planes corresponding to each spin component.}
    \label{fig:Figure3}
\end{figure*}% figure 3

% CrSb
The spin-resolved Fermi surface of CrSb in the relativistic case for $k_z$ = 0.25$\frac{\pi}{c}$ is shown in Fig.~\ref{fig:Figure1}. The $S_x$ and $S_y$ components display planar spin-momentum locking, represented by $Q_{xy}$ and $Q_v$, respectively. The spin-momentum locking of $S_x$ shows two nodal planes in Fig.~\ref{fig:Figure1}(a).
The spin-momentum locking of the $S_y$ component is described by the quadrupole $Q_v$. In the non-relativistic limit or in cubic and orthorhombic symmetries, $Q_v$ has nodal planes along $k_x$ = $\pm$$k_y$. However, in the hexagonal symmetry with relativistic effects, these directions are not high-symmetry lines in the hexagonal Brillouin zone, so $Q_v$ does not exhibit true nodal planes. Therefore, in Fig.~\ref{fig:Figure1}(b), we show two accidental nodal surfaces (ANS), which are remnants of the symmetry-protected nodal planes but are not protected in this case. This behavior arises from the fundamental properties of quadrupoles, already known in classic electromagnetism, which only guarantee nodal lines in the asymptotic limit. More information on these points is reported in Fig. S1 of the supplementary materials. The dominant component $S_z$ exhibits $g$-wave spin-momentum locking consisting of 4 nodal planes, one at $k_z = 0$ and the other 3 nodal planes shown by black solid lines in Fig.~\ref{fig:Figure1}(c). The spectral weight of the subdominant components in the $k$-space is roughly 10\% of the dominant component, although the subdominant components are forbidden in the real space. The RSML of CrSb in the full Brillouin zone is reported in Fig.~\ref{fig:Figure2}. We have demonstrated that only planar spin–momentum locking can survive on the surface of altermagnets\cite{D3NR03681B}. Since the spin–momentum lockings of $S_x$ and S$_y$ are planar, they can be observed on the surface using ARPES without the need for soft X-ray techniques to probe bulk states.

Looking at the entire Table \ref{tab:CrSb_AM}, the RSMLs are represented by a single magnetic quadrupole for all spin components. On top of that, there is the weak ferromagnetism when the N\'eel vector is along the $y$-axis.  Excluding the presence of $Q_0$, the quadrupole table is symmetric with respect to the diagonal. A similar symmetry property is observed in the ferromagnetic phase of the NiAs structure\cite{gong2026relativisticspinmomentumlockingferromagnets}.
Finally, we note that, except at the $\Gamma$ point, there is always at least one nonzero spin–momentum locking throughout the Brillouin zone of CrSb. Therefore, in the relativistic limit, spin splitting occurs at every $\mathbf{k}$-point except at $\Gamma$. This can be understood by noting that the subdominant components produce spin-splitting in the nodal planes of the dominant component.
\begin{table}[h!]
\centering
\begin{tabular}{|c|c|c|c|}
\hline
& \multicolumn{3}{c|}{Spin components} \\
\hline
N\'eel vector & $S_x$ & $S_y$ & $S_z$ \\
\hline
$N \parallel x$ & $Q_{yz}$ & $Q_{xz}$ & $Q_{xy}$ \\
\hline
$N \parallel y$ & $Q_{xz}$ & $Q_{yz}$ & $Q_{0}$ + $Q_{xx}$-$Q_{yy}$\\
\hline
$N \parallel z$ & $Q_{xy}$ & $Q_{xx}$-$Q_{yy}$ & $Q_{yz(3x^2-y^2)}$\\
\hline
\end{tabular}
\caption{Lowest wave symmetry allowed of the RSML of CrSb with NiAs hexagonal structure for the N\'eel vector oriented along the $x$, $y$, $z$ axis and for the different spin components $S_x$, $S_y$, and $S_z$. The diagonal elements represent the spin-momentum locking of the dominant components.}
\label{tab:CrSb_AM}
\end{table}% Table I

\section{Relativistic spin-momentum locking of wurtzite M\lowercase{n}T\lowercase{e}}

% general info about wurtzite  
In this section, we aim to calculate the RSML for the wurtzite structure of MnTe in its altermagnetic phase. The wurtzite crystal structure belongs to the hexagonal space group $P6_3mc$ (no. 186) and is characterized by two independent lattice parameters, $a$ and $c$. More information about the notation used for the real- and momentum-space axes is provided in Figure S2 of the Supplementary Materials. Wurtzite MnTe has been predicted to exhibit high-performance ferroelectricity at elevated temperatures \cite{Bezzerga2025}. In wurtzite MnTe, the preferred orientation of the N\'eel vector lies in the $ab$ plane. Magnetic supercell calculations show that the magnetic ground state of wurtzite MnTe is an antiferromagnetic state with Kramers degeneracy. However, among the wurtzite compounds of the same family, MnTe is the closest to exhibiting altermagnetism.\cite{mavani2025competingantiferromagneticphasesmultiferroic} Therefore, doping, strain, or other external stimuli could favor a transition to the altermagnetic phase.

\begin{table}[h!]
\centering
\begin{tabular}{|c|c|c|c|}
\hline
& \multicolumn{3}{c|}{Spin components} \\
\hline
N\'eel vector & $S_x$ & $S_y$ & $S_z$ \\
\hline
$N \parallel x$ & $Q_{yz}$ + $Q_{y}$ & $Q_{xz}$ + $Q_{x}$ & $Q_{xy}$ \\
\hline
$N \parallel y$ & $Q_{xz}$ + $Q_{y}$ & $Q_{yz}$ + $Q_{x}$ & $Q_{0}$ + $Q_{xx}$-$Q_{yy}$\\
\hline
$N \parallel z$ & $Q_{xy}$ + $Q_{y}$ & $Q_{xx}$-$Q_{yy}$ + $Q_{x}$ & $Q_{yz(3x^2-y^2)}$ \\
\hline
\end{tabular}
\caption{Lowest allowed wave symmetry of the RSML of MnTe with wurtzite hexagonal structure for the N\'eel vector oriented along the $x$, $y$, $z$ axis and for the different spin components $S_x$, $S_y$, and $S_z$. The diagonal elements represent the spin-momentum locking of the dominant components.}
\label{tab:wMnTe_AM}
\end{table}% Table II

% info about wurtzite useful to explain our results
The non-relativistic spin--momentum locking of the altermagnetic phase in the wurtzite structure is the bulk $g$-wave analogue of MnTe in the NiAs structure \cite{D3NR04798A}. The main additional property of the wurtzite structure is the breaking of inversion symmetry, which introduces Rashba spin--momentum locking in the relativistic case. The wurtzite and NiAs crystal structures share the same SML in the absence of spin--orbit coupling; hence, we also expect the same RSML for the quadrupole component. Indeed, our calculations show that the RSML of the wurtzite structure can be obtained by adding the Rashba contribution to the RSML of the NiAs structure. The Rashba contribution to the RSML consists of $Q_{y}$ for $S_x$ and $Q_{x}$ for $S_y$, and it is independent of the N\'eel vector orientation.
% Figure 3
An example of our results is reported in Fig.~\ref{fig:Figure3} for the N\'eel vector along the $x$-axis, where we observe one nodal plane for $S_x$, one nodal plane for $S_y$ and two nodal planes for $S_z$. Together with the results from $k_z = 0$ and $k_z$=-0.25$\frac{\pi}{c}$, it is possible to calculate the multipole expansion.
Therefore, the complete results for the RSML of the wurtzite MnTe are reported in Table~\ref{tab:wMnTe_AM} for all N\'eel vector orientations. We find that the RSML of wurtzite MnTe is identical to that of the NiAs structure shown in Table~\ref{tab:CrSb_AM}, with the addition of the Rashba term.

\begin{figure*}
    \centering
    \includegraphics[width=1\linewidth]{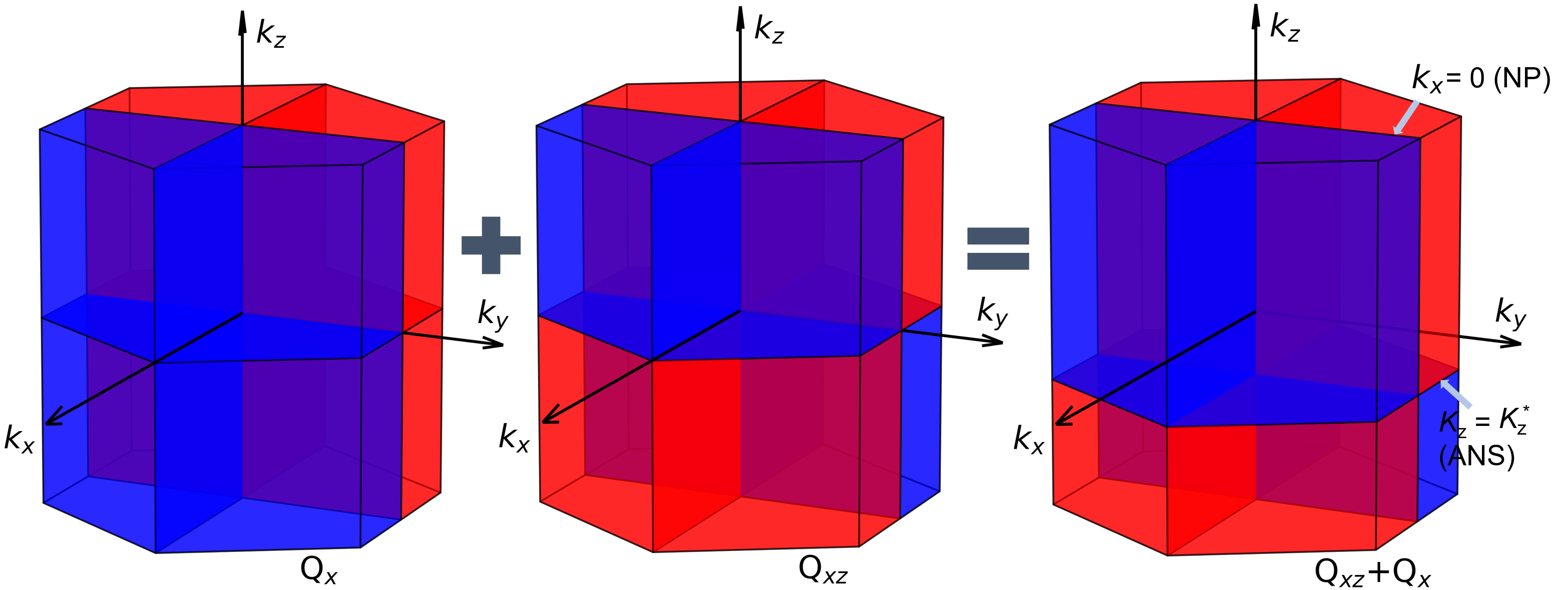}
    \caption{Sum of a dipole ($Q_x$) and a quadrupole ($Q_{xz}$) with a common nodal plane (NP), which is $k_y = 0$. The result is a dipole with the same nodal plane of the addends, an accidental nodal surface (ANS) and zero net magnetization. The ANS is approximately at a constant value of k$_z\approx K_z^* \neq0$. This scenario is realized in the spin-momentum locking of $S_x$, when it is the dominant component.}
    \label{fig:Figure4}
\end{figure*}% Figure 4

\begin{figure*}
    \centering    \includegraphics[width=1\linewidth]{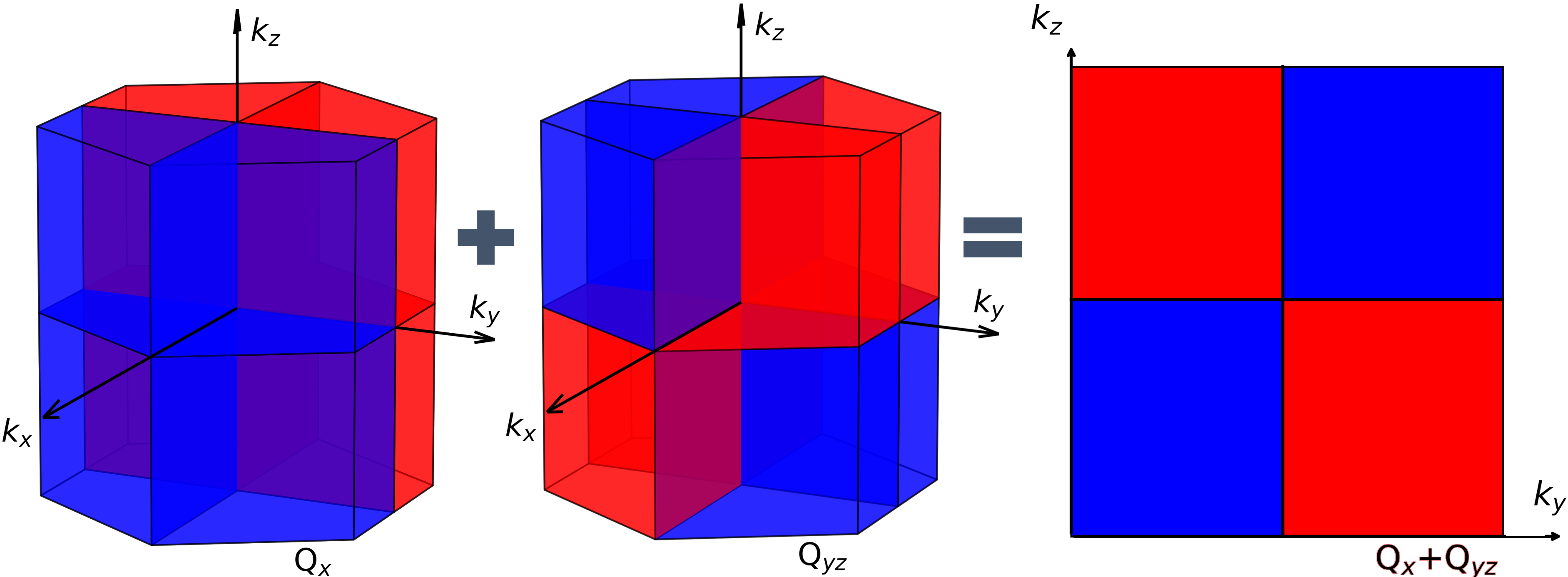}
    \caption{Sum of a dipole ($Q_x$) and a quadrupole ($Q_{yz}$) without a common nodal plane. The sum of them is a complicated SML of which we report only the plane $k_x = 0$, where there are the two nodal lines. The result is a system with two accidental nodal surfaces, two nodal lines (NLs) and zero net magnetization. The nodal lines are at $k_x = k_y = 0$ and $k_x = k_z = 0$. This specific combination appears in the spin-momentum locking of $S_y$, when it is the dominant component.}
    \label{fig:Figure5}
\end{figure*}

% Figure 4 and Figure 5
To better understand the interplay between the Rashba effect and the magnetic quadrupole, we investigated the behavior arising from the combination of a dipole and a quadrupole. Two distinct scenarios can be identified: the first occurs when the dipole and quadrupole share a common nodal plane, while the second occurs when they do not. Examples of the first scenario are the cases $Q_{xz}$+$Q_x$ and $Q_{yz}$+$Q_{y}$.
$Q_{yz}$+$Q_{y}$ is the SML of $S_x$ when it is the dominant component, the analogue $Q_{yz}$+$Q_{y}$ is shown in Fig.~\ref{fig:Figure4}, An example of the second scenario is $Q_{yz}$+$Q_{x}$ which is the SML of $S_y$ when it is the dominant component, shown in Fig.~\ref{fig:Figure5}.
%figure 4
For the first scenario, we plot the magnetic dipole $Q_x$ and quadrupole $Q_{xz}$ in the left and middle panel of Fig.~\ref{fig:Figure4}. If we focus on the region where $k_x > 0$, $Q_x$ appears as a completely blue region, while $Q_{xz}$ is half red and half blue. The sum of the two results in two regions with a predominance of blue, as shown in $Q_{xz} + Q_x$. In the region where $k_x < 0$, the exact opposite happens.
Therefore, by including the breaking of inversion symmetry transforms the magnetic quadrupole with two symmetry-protected nodal planes into a magnetic dipole with one symmetry-protected nodal plane ($k_x = 0$) and one accidental nodal plane ($k_z\approx K_z^* \neq0$), as illustrated in Fig.~\ref{fig:Figure4}.
In centrosymmetric altermagnets, only an even number of nodal planes can exist. When the symmetries protecting these nodal planes are broken---either by structural distortions\cite{ssxp-gz9l,pbbr-hwz4} or by the N\'eel vector breaking rotational symmetry\cite{AutieriRSML}---they transform into accidental nodal surfaces, which also occur in even numbers. To eliminate these accidental nodal planes in centrosymmetric altermagnets, they must merge pairwise as shown in Fig. S17 of the supplementary materials. In contrast, this constraint does not apply to noncentrosymmetric systems, where a single accidental nodal plane can disappear by moving toward the boundary of the Brillouin zone.

%figure 5
In the second scenario, we consider the magnetic quadrupole $Q_{yz}$ and the dipole $Q_x$ and represent the case of $S_y$ dominant component. The quadrupole with two symmetry-protected nodal planes evolves into an object with no symmetry-protected nodal planes, but with two ACNs accompanied by two nodal lines (NLs). Nevertheless, symmetry-protected zero magnetization persists because the two ACNs are related by symmetry, ensuring that the total magnetization remains zero. In the specific case of $Q_{yz}$ and $Q_x$, reported in Fig.~\ref{fig:Figure5}, the NLs are described by the equations $k_x = k_y = 0$ and $k_x = k_z = 0$.

% Figure 5

\begin{figure*}
    \centering
    \includegraphics[width=1\linewidth]{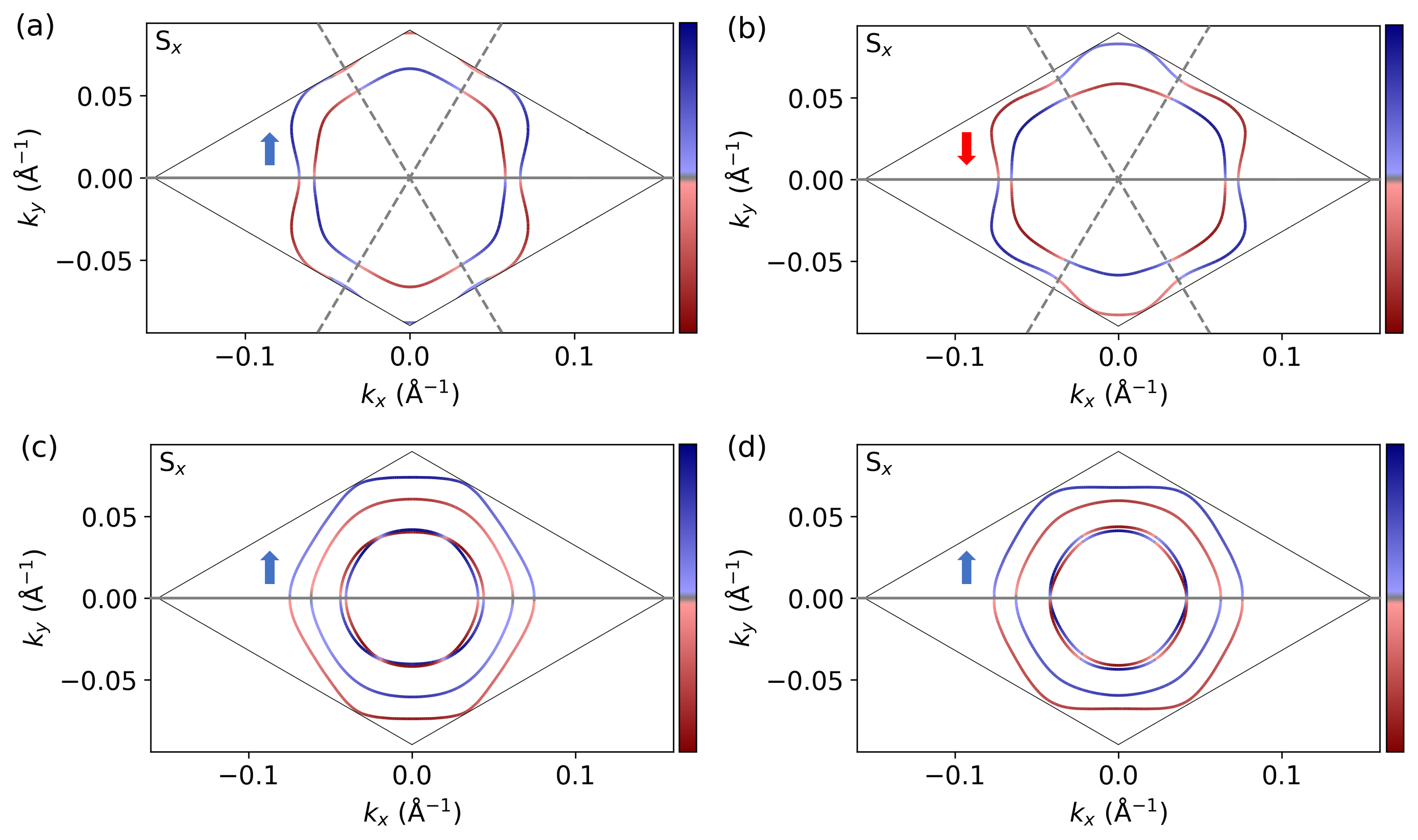}
    \caption{Fermi surfaces for the dominant $S_x$ component at $-1\,\mathrm{eV}$ for (a) $k_z$ = 0.25$\frac{\pi}{c}$ and (b) $k_z$ = -0.25$\frac{\pi}{c}$. With dashed lines, we plot the accidental nodal surfaces that are a reminiscence of the non-relativistic $g$-wave spin-momentum locking. Fermi surfaces for the dominant $S_x$ component at $-2\,\mathrm{eV}$ for (c) $k_z = 0.25\frac{\pi}{c}$ and (d) $k_z$ = -0.25$\frac{\pi}{c}$. Red and blue denote bands with spectral weight of opposite spins. The different behaviour of the spectral weight for the two sets of bands is highlighted by the different behaviors of the arrows simulating the spin directions. Due to the breaking of inversion symmetry, the shapes of Fermi surfaces at $k_z = \pm$0.25$\frac{\pi}{c}$ are different.}
    \label{fig:Figure6}
\end{figure*}

%figure 6
The results reported in Figs.~\ref{fig:Figure4} and~\ref{fig:Figure5} would be visible in the band structure only when the quadrupole and dipole contributions are of comparable magnitude. If one dominates over the other, the band structure displays only the corresponding dominant feature. Wurtzite exhibits a non-relativistic spin splitting that is smaller than that in the NiAs structure; nevertheless, the $d$-wave magnetism remains quite strong. Wurtzite MnTe exhibits strong inversion-symmetry breaking and strong spin--orbit coupling due to the presence of Te atoms. The $p$-wave Rashba term, which arises from the interplay between inversion-symmetry breaking and spin--orbit coupling, is therefore expected to be significant. Hence, both $p$-wave and $d$-wave contributions are strong. To understand which wave dominates, we focus on the case with the N\'eel vector along the $x$-axis for which we calculate the band structure in Fig.~S13.
In Fig.~S13, we show that different bands exhibit distinct features of the $d$-wave, $p$-wave, or a combination of both with the ACN.
At $k_y = 0$ we always have an exact nodal line; we need to check if we also have a nodal line at $k_z = 0$ and see if, upon reversal of $k_z$ from $k_z$ = 0.25$\frac{\pi}{c}$ to $k_z$ = -0.25$\frac{\pi}{c}$, we have a reversal of the spectral weight on the spin-resolved Fermi surface.
We select the band at $-1$ eV, where the even-wave component dominates, and the bands at $-2$ eV, where the $p$-wave component dominates, and report both in Fig.~\ref{fig:Figure6}. For the bands at $-1$ eV shown in Fig.~\ref{fig:Figure6}(a,b), the spin spectral weight on the Fermi surface changes sign upon reversing $k_z$, as highlighted by arrows of different colors representing opposite spins. Therefore, in this energy range, wurtzite MnTe exhibits even-wave magnetism. 
For the bands at $-2$ eV shown in Fig.~\ref{fig:Figure6}(c,d), the spin spectral weight on the Fermi surface does not change sign upon reversing $k_z$, as indicated by arrows of the same color representing identical spins. Therefore, in this energy range, wurtzite MnTe exhibits robust $p$-wave magnetism. Consequently, within this set of bands, wurtzite MnTe behaves as a robust $p$-wave magnet when the N\'eel vector is oriented along the $x$-axis.

%Carmine: We have used different POSCAR notation for CrSb and MnTe, but this did not affect the results :)

\begin{figure*}
    \centering
    \includegraphics[width=1\linewidth]{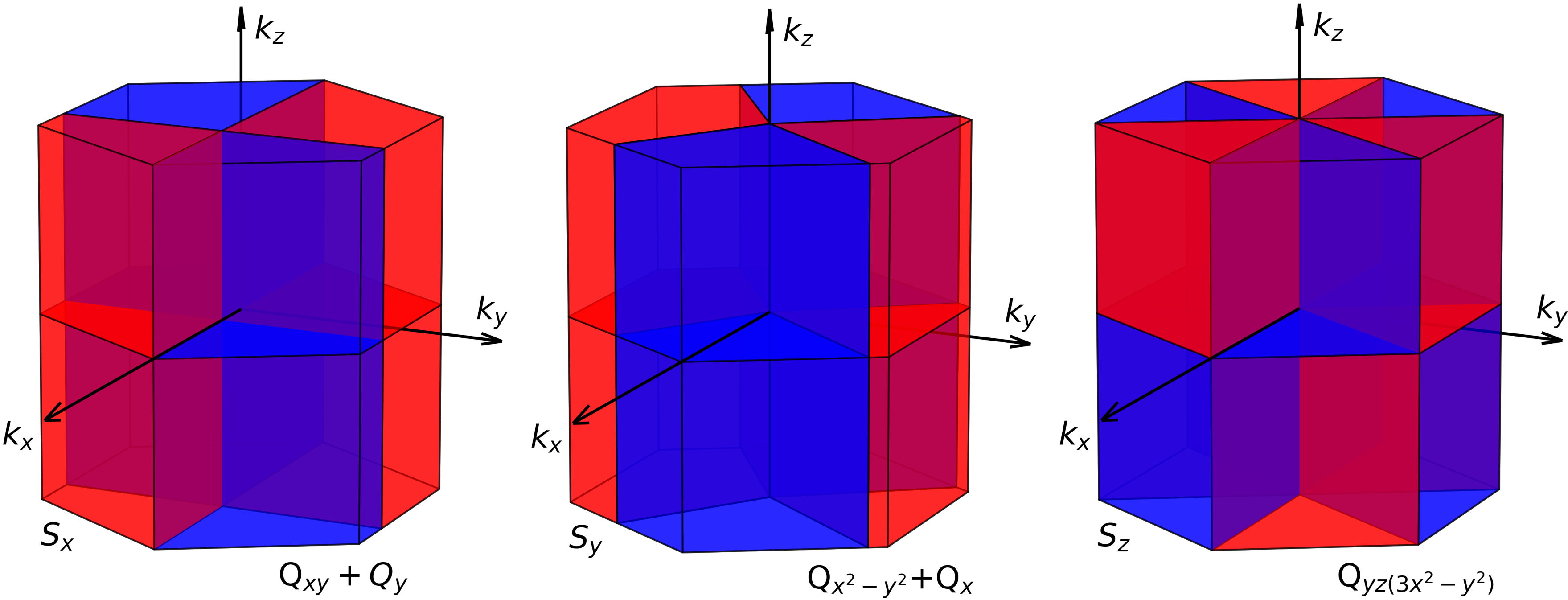}
    \caption{Relativistic spin-momentum locking of wurtzite MnTe with N\'eel vector along the $z$-axis.
    The RSML is composed of $Q_{xy}$ + $Q_{y}$, $Q_{xx}$-$Q_{yy}$ + $Q_{x}$ and  $Q_{yz(3x^2-y^2)}$ for $S_x$, $S_y$ and $S_z$, respectively.
    Red and blue denote regions of the Brillouin zone with opposite spin-splitting. The SML of $S_x$ exhibits 1 nodal plane and 1 accidental nodal surface, the SML of $S_y$ exhibits 2 accidental nodal surfaces and the SML of $S_z$ exhibits 4 nodal planes.}
    \label{fig:Figure7}
\end{figure*}

Once we have clarified the conditions under which different waves can mix and coexist, we can describe our results in terms of the symmetry-allowed lowest wave. With the N\'eel vector along the $x$-axis, the relativistic spin--momentum locking involves $p_y$, $p_x$, and $d_{xy}$ contributions corresponding to the spin components $S_x$, $S_y$, and $S_z$, respectively. Therefore, the dominant spin component $S_x$ exhibits $p_y$ spin--momentum locking, which can be robust for selected bands. With the N\'eel vector aligned along the $y$-axis, both $S_x$ and $S_y$ do not exhibit nodal planes but only ACNs with NLs. The $S_z$ component shows weak ferromagnetism, where the presence of ACNs in the SML of $S_z$ depends on the magnitude of the $s$-wave: for large values of the $s$-wave component, no ACNs are observed, whereas for small values of the $s$-wave component, ACNs can appear, as in MnTe with the NiAs structure.
% N\'eel along z -- Figure 7
The RSML of altermagnetic wurtzite MnTe with N\'eel vector along the $z$-direction is reported in Fig.~\ref{fig:Figure7}. We can observe the coexistence of the $p$-wave for $S_x$, an SML with 2 accidental nodal surfaces for $S_y$ and the $g$-wave for the $S_z$ component. These results are supported by plots of the spin-resolved Fermi surface reported in the Supplementary Materials. Other spin-resolved Fermi surfaces for the in-plane directions of the N\'eel vector and other values of $k_z$ are reported in the supplementary materials in Figs.~S3--S7, while the spin-resolved Fermi surfaces for the N\'eel vector along the $z$-axis are shown in Figs.~S8--S10.
The RSML presented in Fig. \ref{fig:Figure7} would also correspond to the RSML of the (001) surface of CrSb if the surface exhibited a strong Rashba effect.

%These last two sections are completed and almost in a final form
\section{Collinearity in this material class of altermagnet}

Altermagnets are generally non-collinear due to the presence of small cantings induced by spin–orbit coupling \cite{PhysRevB.111.054442,Cheong2024}. Even pure altermagnets can be non-collinear if they host 4 or more magnetic atoms\cite{Fakhredine25b}. However, many of their most relevant properties originate from the collinear component of the magnetic structure, including non-relativistic spin–momentum locking and the anomalous Hall effect. The small deviations from collinearity only quantitatively affect these properties with small quantitative variations, although they can generate weak ferromagnetism or weak ferrimagnetism with a small net magnetic moment. The allowed components of these magnetic moments are connected to the symmetry-allowed components of the anomalous Hall effect. Beyond symmetry analysis, one way to assess the degree of collinearity in density functional theory is to study the relativistic spin-density of states projected onto the three spin components of the two magnetic atoms.

We already reported the spin-canting for MnTe in the NiAs structure\cite{g32j-hnvz}. Repeating the same calculations for wurtzite MnTe, we find that when the N\'eel vector is aligned along the $x$- or $z$-axis, the altermagnetic wurtzite phase remains a pure altermagnet without spin canting, despite spin-orbit and breaking of inversion symmetry. Therefore, we confirm that the RSML is not related to spin canting and is also present in collinear magnetic structures. This holds for both altermagnetic\cite{kq6x-7jfc} and ferromagnetic phases\cite{gong2026relativisticspinmomentumlockingferromagnets}.
In contrast, when the N\'eel vector is oriented along the $y$-axis, the system develops weak ferromagnetism with a net magnetization along the $z$-axis. This state is accompanied by spin canting along the $x$-axis, which can be interpreted as a small rotation of the N\'eel vector in the $xy$ plane. The results are reported in the Supplementary Materials. The oscillatory behaviour exhibited by the $S_z$ component is a signature of the DMI component\cite{PhysRevB.111.054442}. These results on collinearity were the same as those observed in MnTe with the NiAs structure and had the same orders of magnitude \cite{g32j-hnvz}.

 % Figure 7 -- RSML of N\'eel z
% RSML of N\'eel x -- supp. materials S5
% RSML of N\'eel y -- supp. materials S6

% this last part is quite good
\section{Discussion and Conclusions}

In this work, we have investigated the fate of $g$-wave altermagnetism in the presence of relativistic effects. While $g$-wave magnetism is realized in non-relativistic altermagnets, spin--orbit coupling can modify the momentum-space spin structure when the N\'eel vector lies in the $xy$ plane, leading to a reduction of symmetry to $d$-wave. 
% summary section 2
By analyzing CrSb and the altermagnetic phase of wurtzite MnTe with the N\'eel vector oriented along the $z$-axis, we have explicitly evaluated the relativistic spin--momentum locking and identified the symmetry of the resulting spin components in momentum space.
Our results demonstrate that, in both systems, the dominant spin component preserves its $g$-wave character even in the relativistic regime. 
% summary section 3
In contrast, the subdominant components acquire lower angular-momentum symmetries ($d$-wave in CrSb and $p$-wave in wurtzite MnTe) and consequently a lower number of nodal planes. We have compared these findings with previously reported cases, such as MnTe in the NiAs structure with an in-plane N\'eel vector, where relativistic effects reduce the symmetry of the spin components and eliminate the pure $g$-wave character. This comparison clarifies the role of crystal symmetry, spin--orbit coupling, and N\'eel vector orientation in determining whether the $g$-wave structure survives. The Rashba part is independent of the magnetic order (ferromagnetic or altermagnetic) and dominant spin component ($S_x$, $S_y$ or $S_z$).
In the case of wurtzite MnTe, the Rashba $p$-wave term is strong as the even-wave altermagnetic spin-momentum locking. Therefore, there are some bands that are mainly $p$-wave and some bands that are mainly $d$-wave or a combination of them. We provide the recipe to identify bands with robust $p$-wave magnetism in wurtzite MnTe with N\'eel vector along the $x$-axis. The RSML presented for wurtzite MnTe would be the same as that of CrSb under a strong electric field along the z-axis, or as the (001) surface of CrSb if the surface exhibits a strong Rashba effect.
% summary section 4
By examining the spin canting, we find that both material classes are collinear when the N\'eel vector is aligned along the $x$ or $z$ axis, whereas for a N\'eel vector along the $y$ axis, the $S_x$ and $S_z$ spin components exhibit canting. While the $g$-wave in altermagnets appears only in the dominant spin component in some conditions, the corresponding ferromagnetic phase hosts even-wave magnetism only in the subdominant components and therefore does not exhibit $g$-wave magnetism for any component.
% Supp. Materials  -- ferromagnets
In the Supplementary Materials, we have also reported the results for the analogous ferromagnetic phases, where the spin--momentum locking of the dominant component is $s$-wave, while the subdominant components exhibit $d$-wave for the centrosymmetric case and both $p$- and $d$-wave symmetry for the noncentrosymmetric case, but not $g$-wave magnetism.

In centrosymmetric compounds, a given material may exhibit magnetic states with well-defined even-parity components, such as $s$-, $d$-, and $g$-wave symmetries. This behavior is exemplified by systems with the NiAs crystal structure\cite{AutieriRSML}. In contrast, in noncentrosymmetric compounds such as wurztite, the lack of inversion symmetry allows a mixture of even- and odd-parity components, leading to Fermi surfaces with mixed $s$-, $p$-, $d$-, and $g$-wave character, as we demonstrate in this paper.
This motivates a decomposition of the spin-resolved Fermi surface into contributions with definite angular-momentum character using spherical harmonics. Given the spin-resolved Fermi surface at a fixed energy $E_F$ and for a given band $n$ (or Fermi sheet)  and given spin-component $S_i$, one can project the quantity $\mathrm{FS}(\mathbf{k}, E_F, n, S_i)$ onto the basis of spherical harmonics in momentum space, i.e.
\begin{equation}
    \mathrm{FS}(\mathbf{k}, E_F, n, S_i) = \sum_{\ell,m} FS_{\ell m}(E_F, n, S_i)\, Y_{\ell}^{m}(\hat{\mathbf{k}}),
\end{equation}
where $Y_{\ell}^{m}(\hat{\mathbf{k}})$ are the spherical harmonics defined on the spin-resolved Fermi surface and $FS_{\ell m}(E_F,n,S_i)$ are the corresponding projection coefficients. Depending on the symmetries, the presence of nodal planes will constrain to zero the coefficients of some spherical harmonics. For instance, in centrosymmetric systems, the $p$-wave amplitude would have been zero. 
Pure $p$-wave materials would have a large $p$-wave component for every band $n$ and for every Fermi energy E$_F$, while noncentrosymmetric altermagnets\cite{Tenzin2025,leon2025strainenhancedaltermagnetismca3ru2o7} would have, in general, mixed wave components and a large $p$-wave component only for some selected bands and Fermi energies.

\begin{figure*}
    \centering
    \includegraphics[width=1\linewidth]{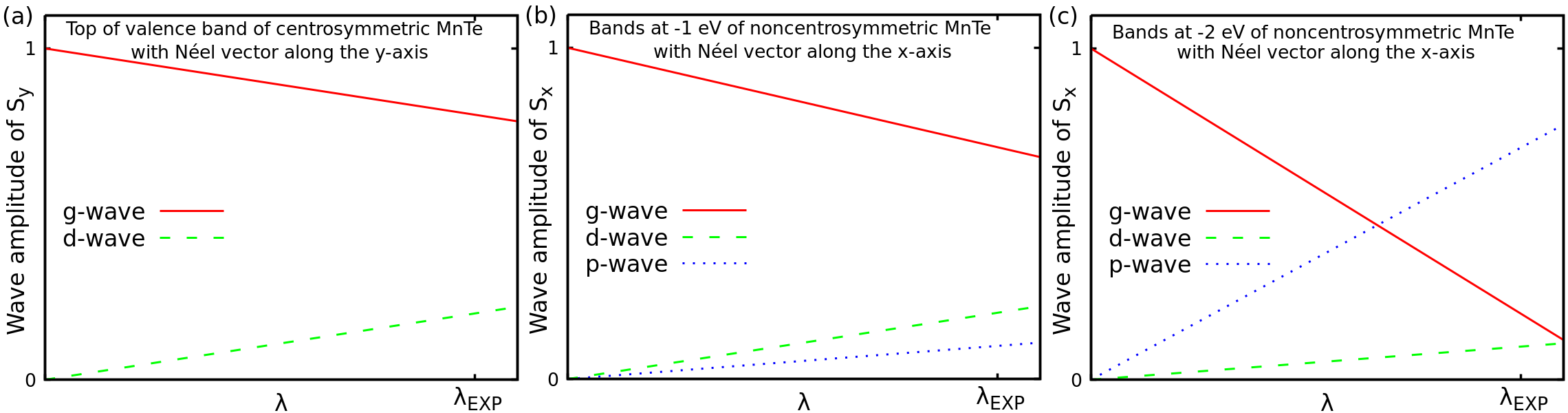}
    \caption{Schematic representation of the mixture of wave amplitudes from a non-relativistic $g$-wave altermagnets and increasing spin-orbit with the N\'eel vector lying in the $xy$ plane. Without spin--orbit coupling, only the $g$-wave component of the spin--momentum locking is symmetry-allowed. (a) Activation of spin--orbit coupling allows the $d$-wave component in the centrosymmetric NiAs structure. (b) Activation of spin--orbit coupling allows the $p$- and $d$-wave components in the noncentrosymmetric wurtzite structure, with the $g$-wave amplitude remaining dominant for the experimental spin--orbit strength. (c) Same as (b) but for another band with the $p$-wave becoming dominant at the experimental spin--orbit coupling.}
    \label{fig:Figure8}
\end{figure*}

%Figure 8
Using these concepts of mixed waves, we present a schematic representation of the behavior of our results: in the previous paper for the dominant $S_y$ component of centrosymmetric MnTe~\cite{AutieriRSML,g32j-hnvz}, and in the present paper for the dominant $S_x$ component of noncentrosymmetric MnTe.
We begin by considering the simplest case, in which the nonrelativistic spin--momentum locking is inherited by the dominant component in the relativistic limit. For both the NiAs and wurtzite structures, with the N\'eel vector oriented out of the plane, the $g$-wave spin--momentum locking for the dominant $S_z$ component remains intact in the presence of spin--orbit coupling, as both $p$- and $d$-wave contributions are forbidden for $S_z$ as discussed in Figs. \ref{fig:Figure2} and \ref{fig:Figure7}. 
The projection of the $S_y$ component of centrosymmetric MnTe is presented in Fig.~\ref{fig:Figure8}(a) as a function of the spin--orbit coupling $\lambda$. When $\lambda = 0$, the $d$-wave component is not allowed and the symmetry is entirely described by $g$-wave spin--momentum locking. Once $\lambda$ is switched on, two nodal planes ($\Gamma$--H$'$ in the literature~\cite{AutieriRSML}) are no longer symmetry-protected, although they do not disappear; therefore, the $g$-wave contribution is still present. At the same time, the $d$-wave component, which is symmetry-allowed, starts to emerge. At the experimental value of the spin--orbit coupling, $\lambda_{\mathrm{EXP}}$, we can assume a still significant $g$-wave contribution from the study of the Fermi surface~\cite{g32j-hnvz}, although the lowest symmetry-allowed component is $d$-wave. Now, let us consider the band at $-1$~eV, where a dominant $S_x$ component of noncentrosymmetric MnTe is observed, as reported in Fig.~\ref{fig:Figure8}(b). Again, when $\lambda = 0$, the Fermi surface is entirely characterized by $g$-wave spin--momentum locking with four nodal planes. Once $\lambda$ is switched on, only the nodal plane at $k_y = 0$ remains symmetry-protected, and both $p$-wave and $d$-wave components become symmetry-allowed. When $\lambda$ reaches $\lambda_{\mathrm{EXP}}$, the system is still predominantly in a $g$-wave state. Finally, let us consider the band at $-2$~eV with a dominant $S_x$ component of noncentrosymmetric MnTe, shown in Fig.~\ref{fig:Figure8}(c). In this case, the $g$-wave contribution decays more rapidly, and a $p$-wave component emerges, leading to a regime characterized by robust $p$-wave behavior at the value of $\lambda_{\mathrm{EXP}}$.

% conclusions

In conclusion, we present two major results. First, we have formulated the symmetry condition under which the dominant spin component retains its $g$-wave character in the relativistic limit. This condition provides a general guideline for identifying altermagnetic materials in which higher-order wave symmetries remain robust against spin--orbit coupling. Second, we identify the cases where the $p$-wave with broken time-reversal symmetry becomes the dominant spin component. From the interplay between different waves, symmetry-protected nodal planes, as well as accidental nodal surfaces and nodal lines, may still emerge. 
We propose a new recipe for realizing $p$-wave magnets by starting from altermagnets with broken inversion symmetry. 
Compared to previous pure odd-wave materials, the main advantage of searching for $p$-wave states in systems with mixed odd- and even-wave components lies in the simpler symmetry requirements. This broadens the range of candidate materials, which may include noncentrosymmetric systems, surfaces of altermagnets\cite{Jeong2025,Jeong2026,lange2026emergentaltermagnetismsurfacesantiferromagnets}, or engineered heterostructures, and also inversion-symmetry breaking induced by external electric fields\cite{PhysRevResearch.2.022025}.

The main drawback, however, is that the $p$-wave character is not robust across all bands, but only appears in a subset of them. Further studies need to address the anisotropy in the response and transport properties\cite{Chakraborty2025} for this family of $p$-wave magnets.
Moreover, our results can be readily generalized to two-dimensional materials, $i$-wave magnets and to systems exhibiting Dresselhaus spin-orbit coupling or persistent spin-helix.

Our investigation also highlights the role of subdominant components, which, even if forbidden or very small in real space, can be of the same order of magnitude as the dominant component in $k$-space. The study of subdominant components is relevant for the spin Hall conductivity\cite{hirakida2025multipoleanalysisspincurrents}, non-linear Hall\cite{mukherjee2025electricfieldcontrolledsecondorder}, Rashba-Edelstein effect\cite{Tenzin2025}, spin photocurrents, and potentially for the interplay between magnetism and superconductivity\cite{Fukaya_2025,Fukaya2025,cv8s-tk4c}. Overall, our analysis establishes a broader framework for understanding relativistic spin textures in altermagnets with higher-order wave symmetries.

After the submission of our paper, the odd-parity was also proposed for the magnons in wurtzite MnTe\cite{du2026oddparitychiralmagnonscollinear}.

\section*{Acknowledgments}

The authors thank T. Wojtowicz, T. Story, J. Sadowski, V. V. Volobuev, P. Barone and G. Cuono for useful discussions. This research was supported by the "MagTop" project (FENG.02.01-IP.05-0028/23) carried out within the "International Research
Agendas" programme of the Foundation for Polish Science, co-financed by the
European Union under the European Funds for Smart Economy 2021-2027 (FENG). C.A. acknowledges support from PNRR MUR project PE0000023-NQSTI. We further acknowledge access to the computing facilities of the Interdisciplinary Center of Modeling at the University of Warsaw, Grant g91-1418, g91-1419, g96-1808, g96-1809 and g103-2540 for the availability of high-performance computing resources and support. We acknowledge the access to the computing facilities of the Poznan Supercomputing and Networking Center, Grants No. pl0267-01, pl0365-01 and pl0471-01.

% To submit to PNAS
\section*{Data, Materials, and Software Availability}
All raw data and analysis scripts supporting the findings of this study have been deposited in Zenodo: [DOI link]. All study data are included in the article and/or SI Appendix.

\section*{Author contributions}
A.F. and C.A. conceptualized and designed the research. X.G. performed the calculations, analyzed the data, and prepared the figures. C.A. acquired funding and supervised the project. All authors contributed to writing, reviewing, and editing the manuscript.

\section*{Competing interests}
The authors declare no competing interests.

\renewcommand{\bibsection}

{\section*{References}}
\bibliographystyle{apsrev4-1}
\FloatBarrier
\bibliography{references}

\end{document}

\title{Supplementary materials: Symmetry-protected nodal planes and accidental nodal \\ surfaces in mixed odd-even wave spin-momentum locking of relativistic altermagnets}

\author{Xujia Gong\orcidA}
\email{xgong@magtop.ifpan.edu.pl}
\affiliation{International Research Centre Magtop, Institute of Physics, Polish Academy of Sciences, Aleja Lotnik\'ow 32/46, PL-02668 Warsaw, Poland}

\author{Amar Fakhredine\orcidF}
\email{amarf@ifpan.edu.pl}
\affiliation{Institute of Physics, Polish Academy of Sciences, Aleja Lotnik\'ow 32/46, 02668 Warsaw, Poland}

\author{Sahar Izadi Vishkayi\orcidS}
\email{izadi.123@gmail.com}
\affiliation{School of Quantum Physics and Matter, Institute for Research in Fundamental Sciences (IPM), P. O. Box 19395-5531, Tehran, Iran}

\author{Carmine Autieri\orcidB}
\email{autieri@magtop.ifpan.edu.pl}
\affiliation{International Research Centre Magtop, Institute of Physics, Polish Academy of Sciences,
Aleja Lotnik\'ow 32/46, PL-02668 Warsaw, Poland}
\affiliation{SPIN-CNR, UOS Salerno, IT-84084 Fisciano (SA), Italy}

\date{\today}

\pacs{}

\maketitle

\section{Symmetry-protected nodal lines and accidental nodal lines for the electric quadrupole Q$_{xx}$-Q$_{yy}$}

In electrostatics, the potential generated by a localized charge distribution can be systematically described using the multipole expansion. For simplicity, we consider a two-dimensional system confined to the $xy$ plane. In this framework, the electric potential is expressed as a hierarchy of terms that encode progressively finer details of the spatial charge distribution. The leading contribution is the monopole moment $Q_{0}$, corresponding to the total charge of the system, which dominates the potential at sufficiently large distances from the source. When the net charge vanishes, the next contribution arises from the dipole moments $Q_{x}$ and $Q_{y}$, representing the components of the electric dipole moment along the $x$ and $y$ directions and describing the first-order spatial separation of positive and negative charges.
Higher-order corrections are determined by the quadrupole moments, which in two dimensions are given by the tensor components $Q_{xx}$, $Q_{yy}$, and $Q_{xy}$. The diagonal elements $Q_{xx}$ and $Q_{yy}$ characterize second-order variations of the charge distribution along the principal axes, while the off-diagonal element $Q_{xy}$ captures the mixed spatial correlations between the $x$ and $y$ directions. These quadrupole terms become particularly relevant when the monopole and dipole moments vanish due to symmetry or overall charge neutrality. Consequently, the multipole moments ${Q_{0}, Q_{x}, Q_{y}, Q_{xx}, Q_{yy}, Q_{xy}}$ provide a systematic characterization of the electrostatic properties of the charge configuration and allow for an accurate approximation of the electric potential in the far-field region.
For a system of $N$ point charges $q_i$ located at positions $(x_i, y_i)$ in the $xy$ plane, the multipole moments reduce to simple sums over the charges:
\begin{equation}\label{monopole}
Q_0 = \sum_{i=1}^{N} q_i
\end{equation}\label{dipole}
\begin{equation}
Q_x = \sum_{i=1}^{N} q_i x_i, \qquad
Q_y = \sum_{i=1}^{N} q_i y_i
\end{equation}
\begin{equation}\label{quadrupole}
Q_{xx} = \sum_{i=1}^{N} q_i x_i^2, \qquad
Q_{yy} = \sum_{i=1}^{N} q_i y_i^2, \qquad
Q_{xy} = \sum_{i=1}^{N} q_i x_i y_i
\end{equation}
The electric potential at a point $(x,y)$ far from all charges ($r \gg |\mathbf{r}_i|$) can then be approximated using the multipole expansion.

\begin{figure}
    \centering
    \includegraphics[width=1\linewidth]{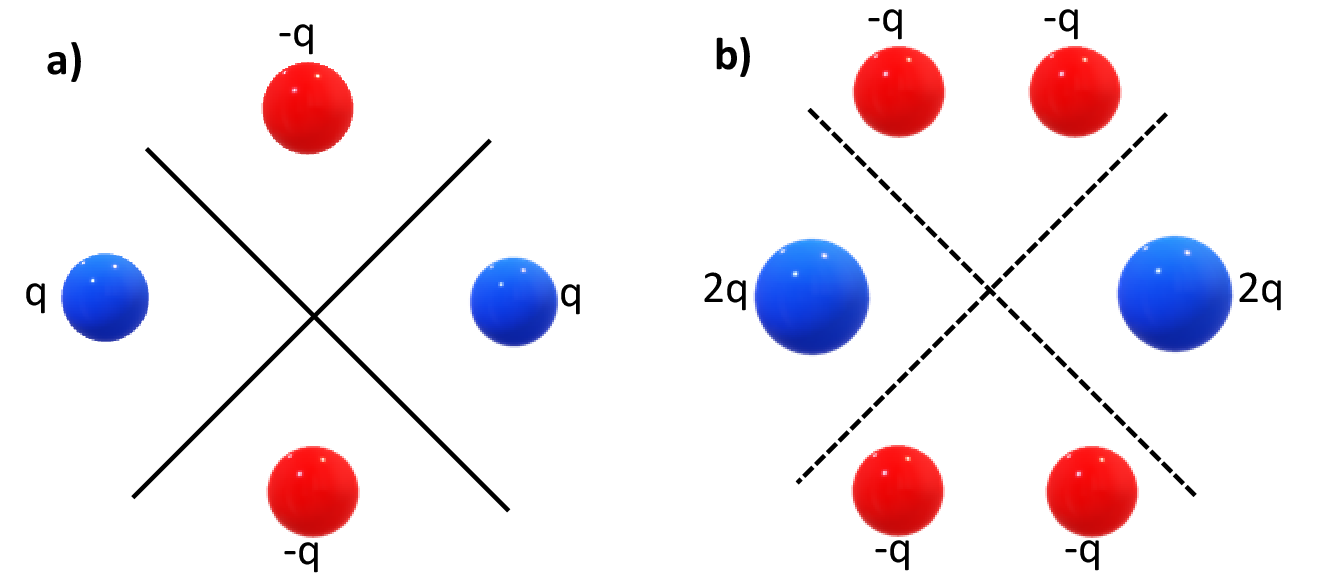}
    \caption{Schematic representation of the electric quadrupole charges in a (a) square symmetrical case, and in a (b) hexagonal symmetrical case. Blue balls represent the positive charges, while red balls represent negative charges. Nodal lines in (a) are represented by solid lines along $x$ = $\pm$ $y$, while the dashed lines in (b) also represent $x$ = $\pm$ $y$, which are not symmetry-protected nodal lines but asymptotic directions where the electric field will be zero at infinite distance from the charges. Both electric configurations are described by the same electric quadrupole moment as the leading term of the multipole expansion. The case in panel (b) is the exact representation of what happened in the $S_y$ component of CrSb.}
    \label{fig:electricquadrupole}
\end{figure}%figure S1

\begin{figure*}
    \centering
    \includegraphics[width=1\linewidth]{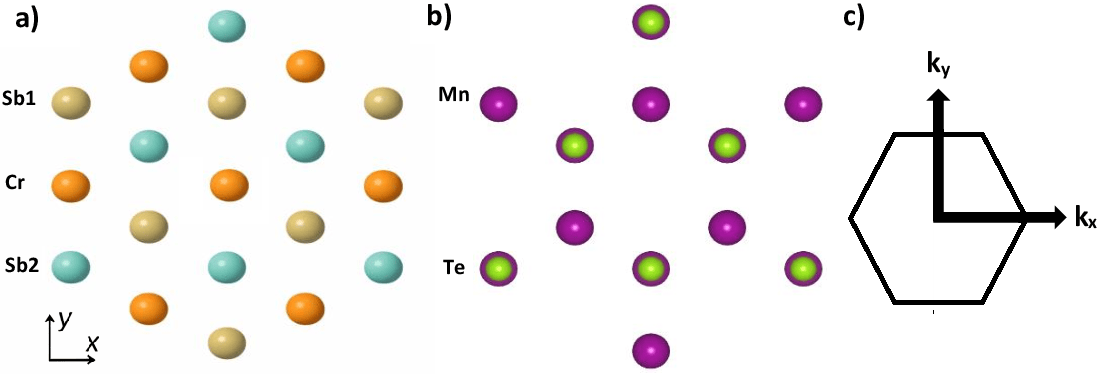}
    \caption{Real-space comparison of the two selected compounds: (a) CrSb (space group No.~194) and (b) wurtzite MnTe (space group No.~186). The $x$ direction is defined along the in-plane lattice vector connecting two nearest-neighbor equivalent atoms, while the $y$ direction is orthogonal to it. The atoms of the same species are, in all cases, on a triangular lattice. (c) Schematic of the reciprocal space in the $k_x$--$k_y$ plane common to both crystal structures, resulting from the definition of the real-space axes.}
    \label{structures}
\end{figure*}%figure S2

In the square-symmetric case, we place four charges: two with charge $+q$ at $(\pm 1,0)$ and two with charge $-q$ at $(0,\pm 1)$. From the formulas of classical electromagnetism, both the electric monopole and dipole moments vanish, while the electric quadrupole moment is $Q_v = Q_{xx} - Q_{yy}$. In this quadrupole, a rotation of $90^\circ$ respect to the origin exchanges $x$ and $y$, reversing the sign of the electric field.
The considered system is shown in Fig.~\ref{fig:electricquadrupole}(a). The electric field exhibits nodal lines along $x=\pm y$, where the electric field is exactly zero. In this case, the nodal lines $x=\pm y$ are represented by solid black lines. Next, let us consider a hexagonal symmetry with six charges: two with charge $+2q$ located at $(\pm 1,0)$ and four with charge $-q$ located at $\left(\pm\frac{1}{2},\pm\frac{\sqrt{3}}{2}\right)$. In this case as well, the electric monopole and dipole moments vanish, while the electric quadrupole moment is a pure $Q_v = Q_{xx} - Q_{yy}$, without any other quadrupole component. The system is shown in Fig.~\ref{fig:electricquadrupole}(b). The electric field does not exhibit nodal lines along $x=\pm y$; however, in the asymptotic limit of large distances from the charges, the electric field vanishes along the lines $x=\pm y$. There are points where the electric field is zero close to the lines $x=\pm y$, corresponding to accidental nodal lines (not symmetry-protected), but the zero-field isocontour does not lie exactly on $x=\pm y$. In this case, the lines $x=\pm y$ are not exact nodal lines and are therefore represented by dashed black lines in Fig.~\ref{fig:electricquadrupole}(b).

In three-dimensional $k$-space, the spin splitting can be systematically expressed in terms of multipole moments. The monopole term $Q_0$ corresponds to a uniform spin splitting, which manifests as ferromagnetism or $s$-wave magnetism. The dipole moments describe $p$-wave magnetic order, such as that arising from the Rashba effect. Finally, the quadrupole moments in three-dimensional $k$-space capture $d$-wave magnetic order, which can be associated with $d$-wave altermagnetism. The formulas (\ref{monopole})–(\ref{quadrupole}) can also be applied using the $k$-space spin splitting obtained from DFT calculations at $k_z$ = 0, 0.25$\frac{\pi}{c}$, and {-0.25}$\frac{\pi}{c}$ to compute the corresponding monopole, dipole, and quadrupole moments, which are reported in Tables I, II, \ref{tab:CrSb_FM}, and \ref{tab:wMnTe_FM}. If only a single multipole is present, explicit calculation is not required, and the multipole can be identified visually. However, the calculation of the dipole using the formulas (\ref{monopole})–(\ref{quadrupole}) and their extension to the three-dimensional space becomes necessary when more multipoles coexist for the same spin components, as in the case of breaking of inversion symmetry. Let us now turn to the case of altermagnets with a quadrupole in three-dimensional $k$-space of the form $Q_v = Q_{xx} - Q_{yy}$. Such a quadrupole can exhibit nodal planes at $k_x=\pm k_y$ if the crystal symmetry is square, whereas nodal planes do not generally appear if the crystal symmetry is hexagonal. In the case of CrSb, which has hexagonal crystal symmetry, the $S_y$ components shown in Fig.~1b of the main text do not exhibit symmetry-protected nodal planes along $k_x=\pm k_y$, but only accidental nodal planes in proximity of $k_x=\pm k_y$. Basically, Fig.~\ref{fig:electricquadrupole}(b) is the classic magnetism analogue of Fig.~1b of the main text, where the spin component$S_y$ on the Fermi surface on the line $k_y$=0 (corresponding to the two charges +2q) has twice and opposite sign of the component$S_y$ on the Fermi surface on the lines $k_y$=$\pm\sqrt{3}$$k_x$ (corresponding to the four charges -q).

\begin{figure*}
    \centering
    \includegraphics[width=1\linewidth]{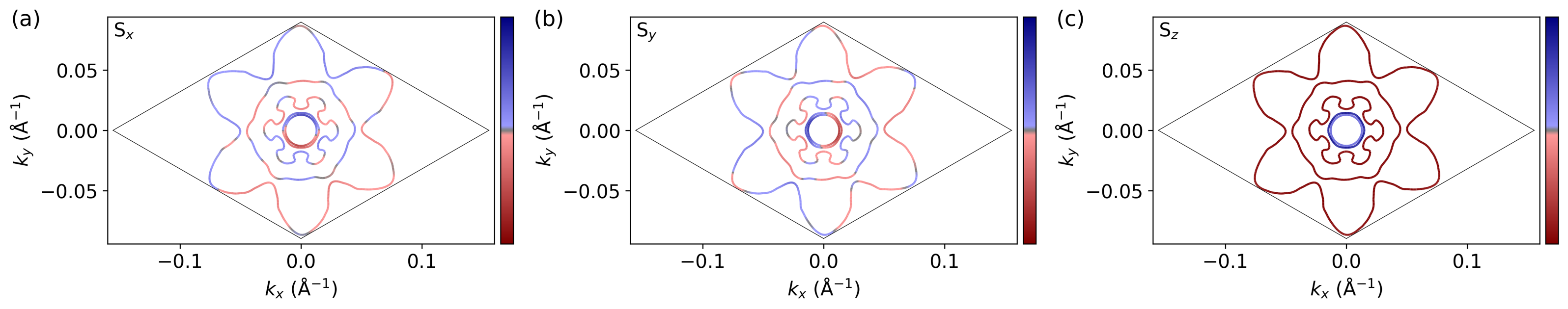}
    \caption{Part of the Brillouin zone in the $k_z$ = 0.25$\frac{\pi}{c}$ plane showing the two-dimensional Fermi surface 2.55 eV above the Fermi surface of the altermagnetic phase of wurtzite MnTe with N\'eel vector along the $x$-axis: (a) the $S_x$ component, (b) the $S_y$ component, and (c) the $S_z$ component. Red and blue denote bands with spectral weight of opposite spins. The black lines denote the nodal plane corresponding to each spin component.}
    \label{fig:S3}
\end{figure*}
\begin{figure*}
    \centering
    \includegraphics[width=1\linewidth]{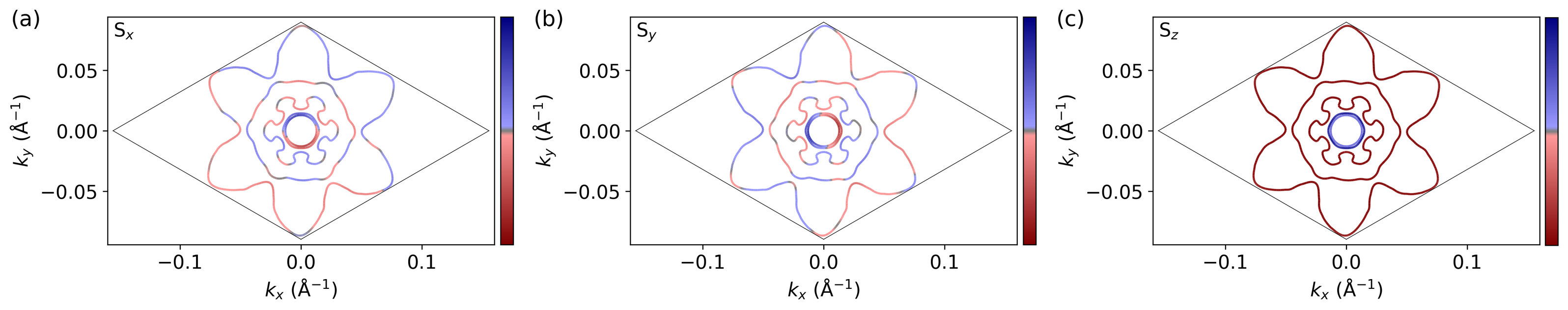}
    \caption{Part of the Brillouin zone in the $k_z$ = -0.25$\frac{\pi}{c}$ plane showing the two-dimensional Fermi surface 2.55 eV above the Fermi surface of the altermagnetic phase of wurtzite MnTe with N\'eel vector along the $x$-axis: (a) the $S_x$ component, (b) the $S_y$ component, and (c) the $S_z$ component. Red and blue denote bands with spectral weight of opposite spins. The black lines denote the nodal plane corresponding to each spin component.}
    \label{fig:S4}
\end{figure*}

\begin{figure*}
    \centering
    \includegraphics[width=1\linewidth]{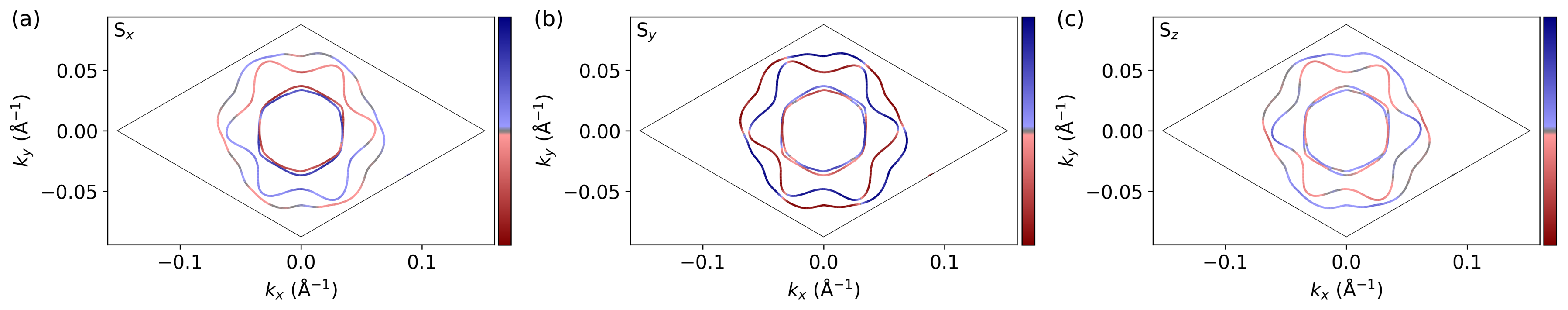}
    \caption{Part of the Brillouin zone in the $k_z$ = 0.25$\frac{\pi}{c}$ plane showing the two-dimensional Fermi surface 2.55 eV above the Fermi surface of the altermagnetic phase of wurtzite MnTe with N\'eel vector along the $y$-axis: (a) the $S_x$ component, (b) the $S_y$ component, and (c) the $S_z$ component. Red and blue denote bands with spectral weight of opposite spins. No nodal planes are present in this case.}
    \label{fig:y_kz25_wmnte}
\end{figure*}

\begin{figure*}
    \centering
    \includegraphics[width=1\linewidth]{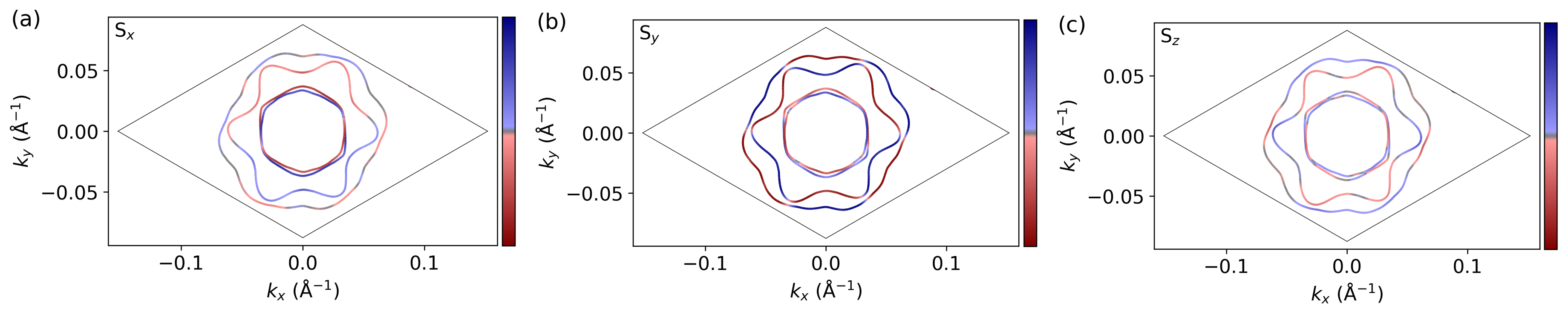}
    \caption{Part of the Brillouin zone in the $k_z$ = -0.25$\frac{\pi}{c}$ plane showing the two-dimensional Fermi surface 2.55 eV above the Fermi surface of the altermagnetic phase of wurtzite MnTe with N\'eel vector along the $y$-axis: (a) the $S_x$ component, (b) the $S_y$ component, and (c) the $S_z$ component. Red and blue denote bands with spectral weight of opposite spins. No nodal planes are present in this case.}
    \label{fig:y_kz-25_wmnte}
\end{figure*}

\begin{figure*}
    \centering
    \includegraphics[width=1\linewidth]{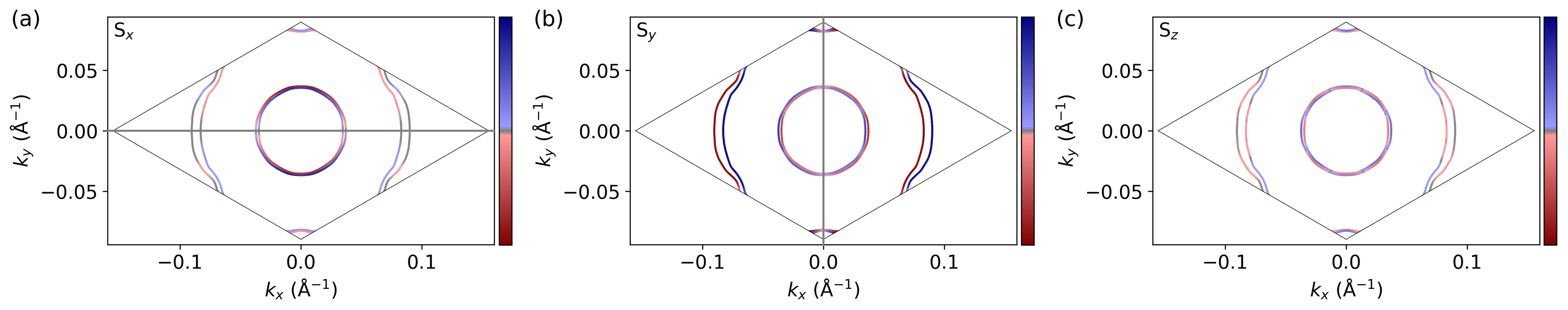}
    \caption{Part of the Brillouin zone in the $k_z = 0$ plane showing the two-dimensional Fermi surface 2.55 eV above the Fermi surface of the altermagnetic phase of wurtzite MnTe with N\'eel vector along the $y$-axis: (a) the $S_x$ component, (b) the $S_y$ component, and (c) the $S_z$ component. Red and blue denote bands with spectral weight of opposite spins. The black lines denote the nodal lines corresponding to each spin component.}
    \label{fig:y_kz0_wmnte}
\end{figure*}%S7

\begin{figure*}
    \centering
    \includegraphics[width=1\linewidth]{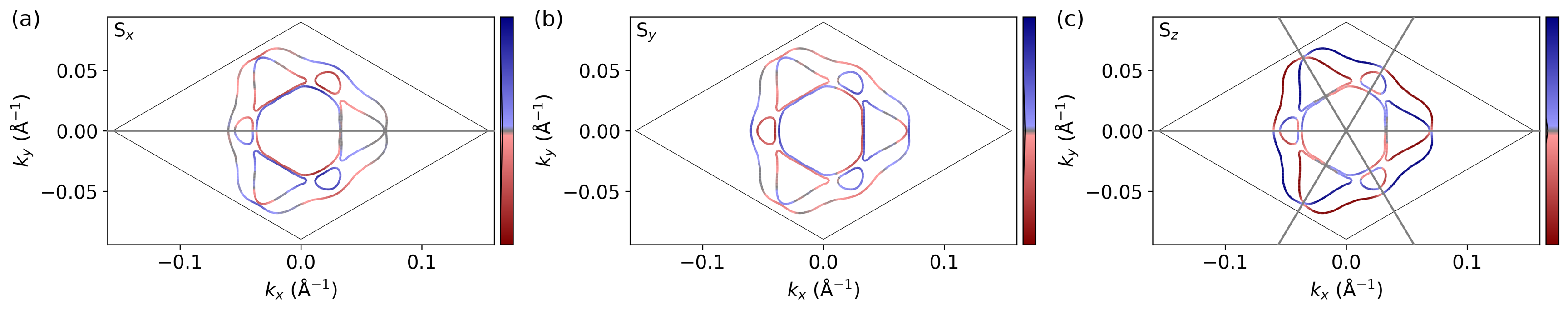}
    \caption{Part of the Brillouin zone in the $k_z$ = 0.25$\frac{\pi}{c}$ plane showing the two-dimensional Fermi surface 2.40 eV above the Fermi surface of the altermagnetic phase of wurtzite MnTe with N\'eel vector along the $z$-axis: (a) the $S_x$ component, (b) the $S_y$ component, and (c) the $S_z$ component. Red and blue denote bands with spectral weight of opposite spins. The black lines denote the nodal plane corresponding to each spin component.}
    \label{fig:z_kz25_wmnte}
\end{figure*}%S8

\begin{figure*}
    \centering
    \includegraphics[width=1\linewidth]{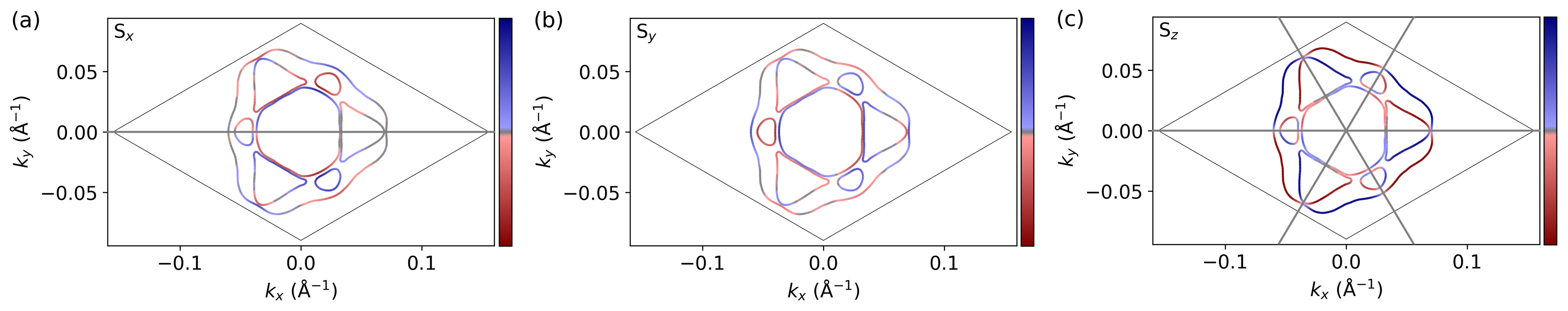}
    \caption{Part of the Brillouin zone in the $k_z$ = -0.25$\frac{\pi}{c}$ plane showing the two-dimensional Fermi surface 2.40 eV above the Fermi surface of the altermagnetic phase of wurtzite MnTe with N\'eel vector along the $z$-axis: (a) the $S_x$ component, (b) the $S_y$ component, and (c) the $S_z$ component. Red and blue denote bands with spectral weight of opposite spins. The black lines denote the nodal plane corresponding to each spin component.}
    \label{fig:z_kz-25_wmnte}
\end{figure*}%S9

\begin{figure*}
    \centering
    \includegraphics[width=1\linewidth]{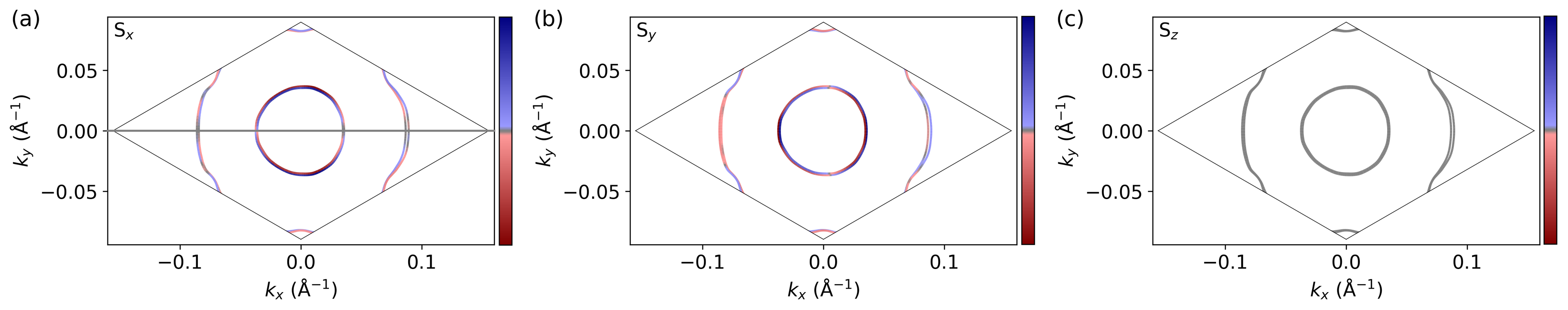}
    \caption{Part of the Brillouin zone in the $k_z = 0$ plane showing the two-dimensional Fermi surface 2.40 eV above the Fermi surface of the altermagnetic phase of wurtzite MnTe with N\'eel vector along the $z$-axis: (a) the $S_x$ component, (b) the $S_y$ component, and (c) the $S_z$ component. Red and blue denote bands with spectral weight of opposite spins. The black lines denote the nodal plane corresponding to each spin component. The plane $k_z$=0 is a nodal plane for the$S_z$ component.}
    \label{fig:z_kz0_wmnte}
\end{figure*}%S10

\begin{figure*}
    \centering
    \includegraphics[width=1\linewidth]{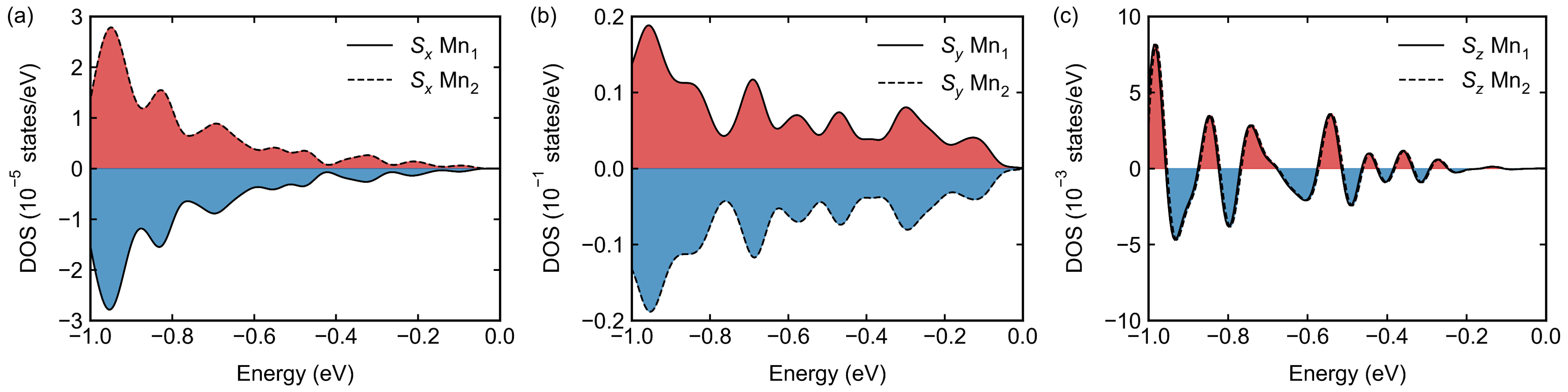}
    \caption{Relativistic spin density of states in the energy range from $-1$ to $0$ eV for wurtzite MnTe, calculated for the N\'eel vector aligned along the $y$-axes. (a) $S_x$ is responsible for the slight rotation of the N\'eel vector within the $xy$ plane. (b) $S_y$ represents the dominant component of the N\'eel vector. (c) The $S_z$ component gives rise to weak ferromagnetism along the $z$-axis.}
\label{fig:S11}
\end{figure*}

\begin{figure*}
    \centering
    \includegraphics[width=1\linewidth]{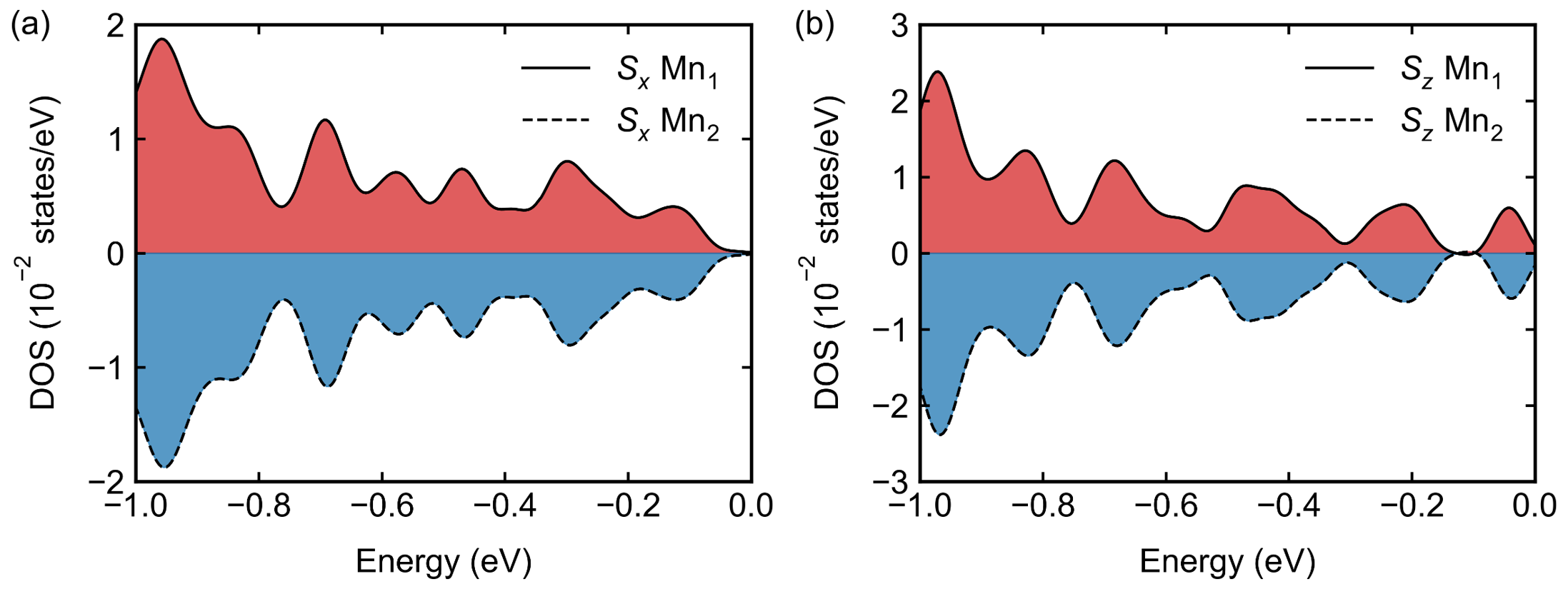}
    \caption{Relativistic spin density of states in the energy range from $-1$ to $0$ eV for wurtzite MnTe. (a) $S_x$ represents the dominant component calculated for the N\'eel vector aligned along the $x$-axis. (b)  $S_z$ represents the dominant component calculated for the N\'eel vector aligned along the $z$-axis.}
    \label{fig:S12}
\end{figure*}

\section{Computational details and crystal structure}

First-principles calculations were performed using the Vienna Ab~initio Simulation Package (VASP)~\cite{kresse1993ab,kresse1996efficiency}, 
within the framework of density functional theory (DFT) and employing the projector augmented wave (PAW) method~\cite{kresse1999ultrasoft}. 
The exchange--correlation potential was described using the generalized gradient approximation (GGA) with the Perdew--Burke--Ernzerhof (PBE) functional~\cite{perdew1996generalized}. 
A plane-wave cutoff energy of 400~eV was used throughout, and the total energy convergence criterion was set to $10^{-5}$~eV. 
To account for electron--electron correlations among localized $d$ orbitals, the DFT+$U$ formalism~\cite{liechtenstein1995density} was adopted. 
The values used for the Columb repulsion are $U = 2$~eV for Cr and $U = 4$~eV for Mn, with Hund’s exchange coupling set to $J_H = 0.15U$. A Monkhorst--Pack $k$-point grid of $15 \times 15 \times 7$ was used for spin-resolved band structure calculations, 
while a denser mesh of $15 \times 15 \times 8$ for CrSb and $18 \times 18 \times 8$ for MnTe were adopted for constructing the two-dimensional (2D) Fermi surface.
The calculated Fermi surface accuracy was set to 0.022~\AA$^{-2}$ for CrSb and 0.016~\AA$^{-2}$ for MnTe. In all 2D Fermi surfaces, the zero of the spin component is represented by the gray color.

In this paper, we have selected CrSb from space group 194 and wurtzite MnTe from space group 186. The real space comparison between the two space groups is reported in Fig.~\ref{structures}. The x-direction is defined as the direction of the in-plane lattice vector, while the y-direction is orthogonal to it.

\section{Fermi surfaces of the altermagnetic phase of M\lowercase{n}T\lowercase{e} in the wurtzite phase}

To analyze the relativistic spin--momentum locking in a system with more than one multipole, we plot the spin-resolved Fermi surfaces at $k_z = 0$, $0.25\frac{\pi}{c}$, and $-0.25\frac{\pi}{c}$. 
For the N\'eel vector aligned along the $x$-axis, these spin-resolved Fermi surfaces are shown in Fig.~3 of the main text and in Figs.~S3--S4. In this case, there is one nodal plane for $S_x$, one nodal plane for $S_y$, and two nodal planes for $S_z$.
For the N\'eel vector along the $y$-axis, the corresponding spin-resolved Fermi surfaces are reported in Figs.~S5--S7. In this configuration, there are no nodal planes for any spin component. However, at $k_z=0$ in Fig.~S7, we highlight one nodal line for $S_x$ and one nodal line for $S_y$.
For the N\'eel vector along the $z$-axis, the corresponding Fermi surfaces are shown in Figs.~S8--S10. In this case, there is one nodal plane for $S_x$, four nodal planes for $S_z$, and no nodal planes for $S_y$.
We report only the symmetry-protected nodal planes and nodal lines visible in the plots, without plotting the accidental nodal surfaces.

%\begin{figure*}
%    \centering
%    \includegraphics[width=1\linewidth]{x_kz0_fermi2d_CrSb_z.png}
%    \caption{Relativistic spin-momentum locking of wurtzite MnTe with N\'eel vector along the $x$-axis. Red and blue denote regions of the Brillouin zone with opposite spin-splitting. The SML of$S_x$ exhibits 1 nodal plane and 1 accidental nodal surface, the SML of$S_y$ exhibits 1 nodal plane and 1 accidental nodal surface and the SML of $ S_z$ exhibits 2 nodal planes.}
%    \label{fig:S5}
%\end{figure*} % RSML of Neel x -- supp. materials

%\begin{figure*}
%    \centering
%    \includegraphics[width=1\linewidth]{x_kz0_fermi2d_CrSb_z.png}
%    \caption{Relativistic spin-momentum locking of wurtzite MnTe with N\'eel vector along the $y$-axis. Red and blue denote regions of the Brillouin zone with opposite spin-splitting. The SMLs of$S_x$,$S_y$, and $ S_z$ exhibit 2 accidental nodal surfaces.}
%    \label{fig:S6}
%\end{figure*} % RSML of Neel y -- supp. materials

\begin{figure*}
    \centering
    \includegraphics[width=1\linewidth]{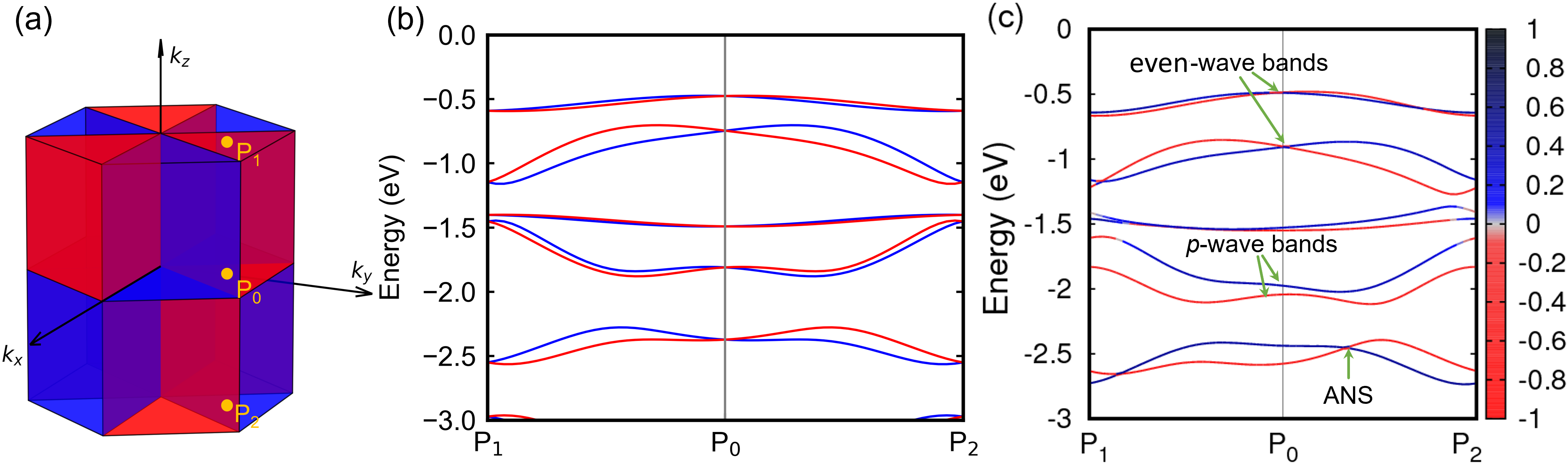}
    \caption{
(a) Brillouin zone of wurtzite MnTe with $g$-wave non-relativistic spin-momentum locking, showing the positions of the $k$-points P$_1$, P$_0$, and P$_2$. (b) Band structure of wurtzite MnTe with magnetization $\parallel x$ without SOC along the $k$-path P$_1$-P$_0$-P$_2$. (c) $S_x$ component spin-resolved band structure of wurtzite MnTe with magnetization $\parallel x$ along the $k$-path P$_1$-P$_0$-P$_2$, where red and blue lines represent spin-up and spin-down states, respectively.}
    \label{fig:bz_bs_wmnte_neelx}
\end{figure*}

% section S4
\section{Collinearity of the altermagnetic phase of M\lowercase{n}T\lowercase{e} in the wurtzite phase}

We report the relativistic density of states (DOS) for wurtzite MnTe. The system has two Mn atoms in the unit cell, which we denote as Mn$_1$ and Mn$_2$. 
We first focus on the N\'eel vector along the $y$-direction. Figures~S11(a,b,c) show the relativistic DOS of the spin components $S_x$, $S_y$, and $S_z$, respectively. The DOS of the dominant component $S_y$ has opposite contributions on the atoms Mn$_1$ and Mn$_2$, indicating that these atoms exhibit opposite $S_y$ components. The same behavior occurs for the $S_x$ component; however, its magnitude is about three orders of magnitude smaller. This implies a very small rotation of the N\'eel vector in the $xy$ plane. The DOS of the $S_z$ component shows contributions with the same sign on Mn$_1$ and Mn$_2$, which generates weak ferromagnetism. 
Figures~S12(a) and S12(b) show the relativistic DOS of the main spin component of the pure altermagnetic phases for the N\'eel vector aligned along the $x$- and $z$-axes. The relativistic DOS of the main component for the $x$- and $y$-axes are very similar, whereas a noticeable difference appears for the $z$-axis. This behavior is consistent with the magnetocrystalline anisotropy reported for the $\mathcal{PT}$-symmetric phase (antiferromagnetic phase with Kramers degeneracy) of wurtzite MnTe, where it was shown that the energies of magnetic states with spins in the $ab$ plane are nearly degenerate, while a relatively strong energy variation occurs for spins oriented along the $z$-axis~\cite{mavani2025competingantiferromagneticphasesmultiferroic}.

\begin{figure*}
    \centering
    \includegraphics[width=1\linewidth]{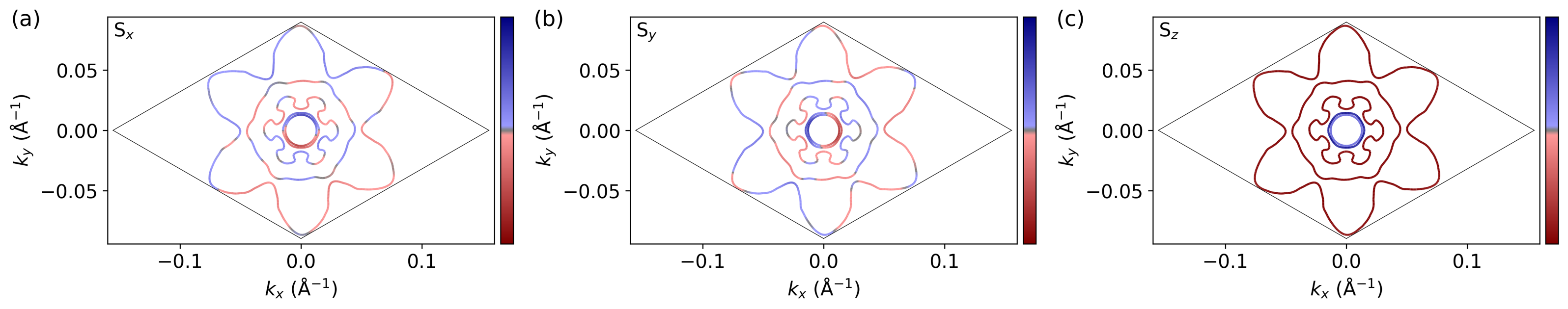}
    \caption{Part of the Brillouin zone in the $k_z$ = 0.25$\frac{\pi}{c}$ plane showing the two-dimensional Fermi surface 2 eV above the Fermi surface of ferromagnetic phase of wurtzite MnTe with N\'eel vector along the $z$-axis: (a) the $S_x$ component, (b) the $S_y$ component, and (c) the $S_z$ component. Red and blue denote bands with spectral weight of opposite spins.}
    \label{fig:wmente_fm_z_fermi2d_kz25}
\end{figure*}%S14

\begin{figure*}
    \centering
    \includegraphics[width=1\linewidth]{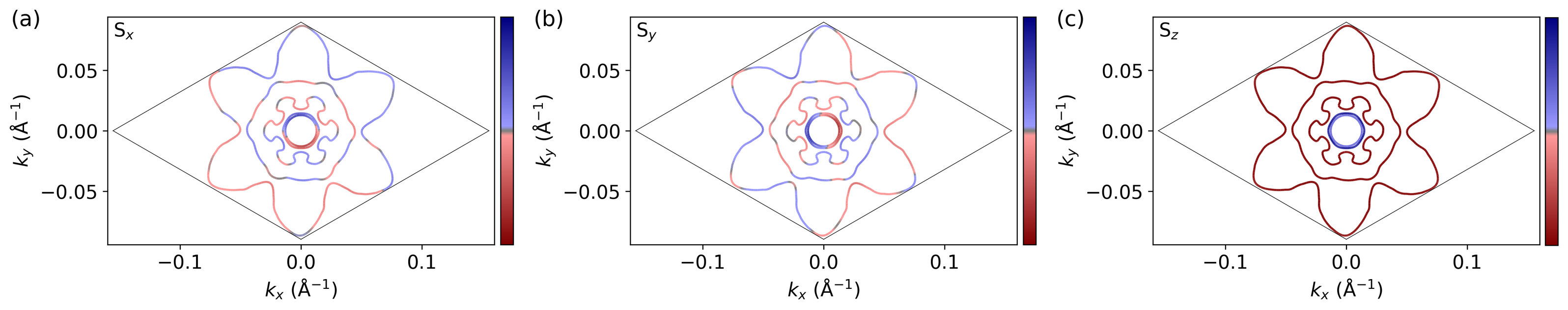}
    \caption{Part of the Brillouin zone in the $k_z$ = -0.25$\frac{\pi}{c}$ plane showing the two-dimensional Fermi surface 2 eV above the Fermi surface of ferromagnetic phase of wurtzite MnTe with N\'eel vector along the $z$-axis: (a) the $S_x$ component, (b) the $S_y$ component, and (c) the $S_z$ component. Red and blue denote bands with spectral weight of opposite spins.}
    \label{fig:wmente_fm_z_fermi2d_kz-25}
\end{figure*}%S15

\begin{figure*}
    \centering
    \includegraphics[width=1\linewidth]{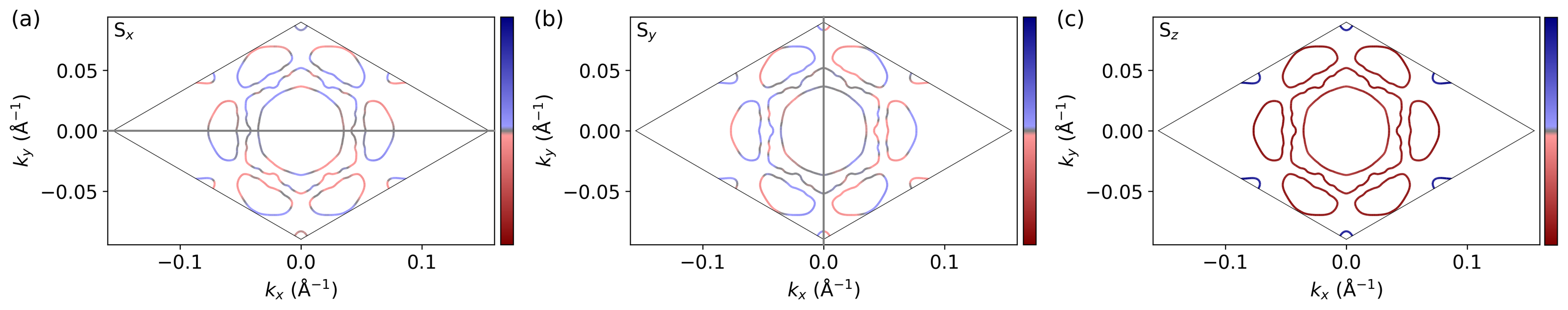}
    \caption{Part of the Brillouin zone in the $k_z = 0$ plane showing the two-dimensional Fermi surface 2.30 eV above the Fermi surface of ferromagnetic phase of wurtzite MnTe with N\'eel vector along the $z$-axis: (a) the $S_x$ component, (b) the $S_y$ component, and (c) the $S_z$ component. Red and blue denote bands with spectral weight of opposite spins. The black lines denote the nodal lines for the $S_x$ and $S_y$ components.}
    \label{fig:wmente_fm_z_fermi2d_kz0}
\end{figure*}%S16

\section{Accidental nodal surface}

We performed band structure calculations for wurtzite MnTe with the N\'eel vector aligned along the $x$-axis, considering the $k$-path from $P_1$ to $P_2$ passing through $P_0$, both without and with spin--orbit coupling. The $k$-points $P_1$, $P_0$, and $P_2$ are defined in direct coordinates as $(0, 1/2, 0.5)$, $(0, 1/2, 0)$, and $(0, 1/2, -0.5)$, respectively, as shown in Fig.~S13(a) where there is the non-relativistic $g$-wave spin-momentum locking. 
This choice of the $k$-path enables us to plot the band structure in the plane $k_x = 0$ with  $k_y \neq 0$, which is a nodal plane only for the spin components $S_y$ and $S_z$; consequently, only the dominant component $S_x$ remains. The $k$-path $P_1$--$P_0$--$P_2$ intersects the nodal plane of non-relativistic spin--momentum locking $k_z = 0$ at the point $P_0$.
In Fig.~S13(b), we present the non-relativistic band structure in the energy range from the top of the valence band down to $-3$~eV. In the absence of SOC, we observe a non-relativistic spin splitting that is odd with respect to $P_0$, where a double degeneracy between spin-up and spin-down states is preserved.
In Fig.~S13(c), we show the relativistic band structure, focusing on the dominant $S_x$ component, within the same energy window. When SOC is included, Rashba-type interactions emerge, leading to different behaviors depending on the spin-momentum locking of the bands. If no splitting occurs at the $\Gamma$ point, the bands retain predominantly even-wave (mixed $g$ and $d$-wave) spin-momentum locking; otherwise, they exhibit increased $p$-wave spin-momentum locking or develop accidental nodal surfaces (ANS). For weak Rashba coupling, the bands largely preserve their $d$-wave, as observed for the states near the top of the valence band. In contrast, strong Rashba coupling leads to fully spin-polarized bands, as seen for a band around $-2$~eV, which is predominantly $p$-wave. When the two contributions are comparable, an accidental nodal surface can form, as observed for the band near $-2.5$~eV. An arrow highlights the location of the ANS. Since the ANS is not symmetry-protected, its position in $k$-space can vary slightly between different bands.
This analysis enables us to identify bands with dominant $p$-wave spin-momentum locking, which are subsequently used to construct the Fermi surface shown in Fig.~6 of the main text.

%%%%%%%%%%%%%%%%%%%%%%%%%%%%%%%%%%%%%%%%%%%%%%%%%
%%%%%%%%%%%%%%%%%%%%%%%%%%%%%%%%%%%%%%%%%%%%%%%%%

\begin{table}[h!]
\centering
\begin{tabular}{|c|c|c|c|}
\hline
& \multicolumn{3}{|c|}{Spin components} \\
\hline
Magnetization direction &$S_x$ &$S_y$ &$S_z$ \\
\hline
$M \parallel x$ & $Q_0$ & $Q_{xy}$ & $Q_{xz}$ \\
\hline
$M \parallel y$ & $Q_{xy}$ & $Q_0$ &  $Q_{yz}$ \\
\hline
$M \parallel z$ & $Q_{xz}$ & $Q_{yz}$  & $Q_0$ \\
\hline
\end{tabular}
\caption{
Lowest wave symmetry allowed of the RSML of ferromagnetic MnTe with NiAs hexagonal structure for the magnetization (M) oriented along the $x$, $y$, $z$ axis and for the different spin components $S_x$, $S_y$, and $S_z$.}
\label{tab:CrSb_FM}
\end{table} % Table I -- supp materials

\begin{table}[h!]
\centering
\begin{tabular}{|c|c|c|c|}
\hline
& \multicolumn{3}{c|}{Spin components} \\
\hline
Magnetization direction &$S_x$ &$S_y$ &$S_z$ \\
\hline
$M \parallel x$ & $Q_0$ + $Q_y$ & $Q_{xy}$ + $Q_x$ & $Q_{xz}$ \\
\hline
$M \parallel y$ & $Q_{xy}$ + $Q_y$ & $Q_0$ + $Q_x$ &  $Q_{yz}$ \\
\hline
$M \parallel z$ & $Q_{xz}$ + $Q_y$ & $Q_{yz}$ + $Q_x$ & $Q_0$ \\
\hline
\end{tabular}
\caption{Lowest wave symmetry allowed of the RSML of ferromagnetic MnTe with wurtzite hexagonal structure for the magnetization (M) oriented along the $x$, $y$, $z$ axis and for the different spin components $S_x$, $S_y$, and $S_z$.}
\label{tab:wMnTe_FM}
\end{table} % Table II - supp. materials

\section{Ferromagnetic phases}

In the altermagnetic phase, we found that the dominant spin component is $g$-wave when the N\'eel vector is along the $z$-axis for both NiAs and wurtzite crystal structure. Therefore, in the ferromagnetic phases of both NiAs and wurtzite crystal structure, where the dominant component is the $s$-wave, we expect the $g$-wave contribution to be absent. This expectation is confirmed by our spin-resolved Fermi surface calculations. We report the results for the ferromagnetic phase of NiAs in Table~\ref{tab:CrSb_FM}, which were already reported by us in another paper. For the ferromagnetic wurtzite, the results are summarized in Table~\ref{tab:wMnTe_FM}. In both cases, we find no $g$-wave.
As in the case of altermagnetic phases, the relativistic spin--momentum locking (RSML) of the wurtzite structure in ferromagnetic phases can be obtained by combining the RSML of the NiAs structure with an additional Rashba contribution. 
The spin-resolved Fermi surfaces for ferromagnetic wurtzite with magnetization oriented along the $z$-axis are shown in Figs.~S14--S16 for $k_z$ = 0.25$\frac{\pi}{c}$, $k_z$ = -0.25$\frac{\pi}{c}$, and $k_z = 0$, respectively. The Fermi surface has a hexagonal symmetry and the dominant component is $S_z$. The $S_x$ and $S_y$ components do not exhibit any nodal planes, as can be seen from Figs.~S14 and S15. Instead, they host only nodal lines, which are visible in Figs.~S16(a) and S16(b).

\begin{figure*}
    \centering
    \includegraphics[width=1\linewidth]{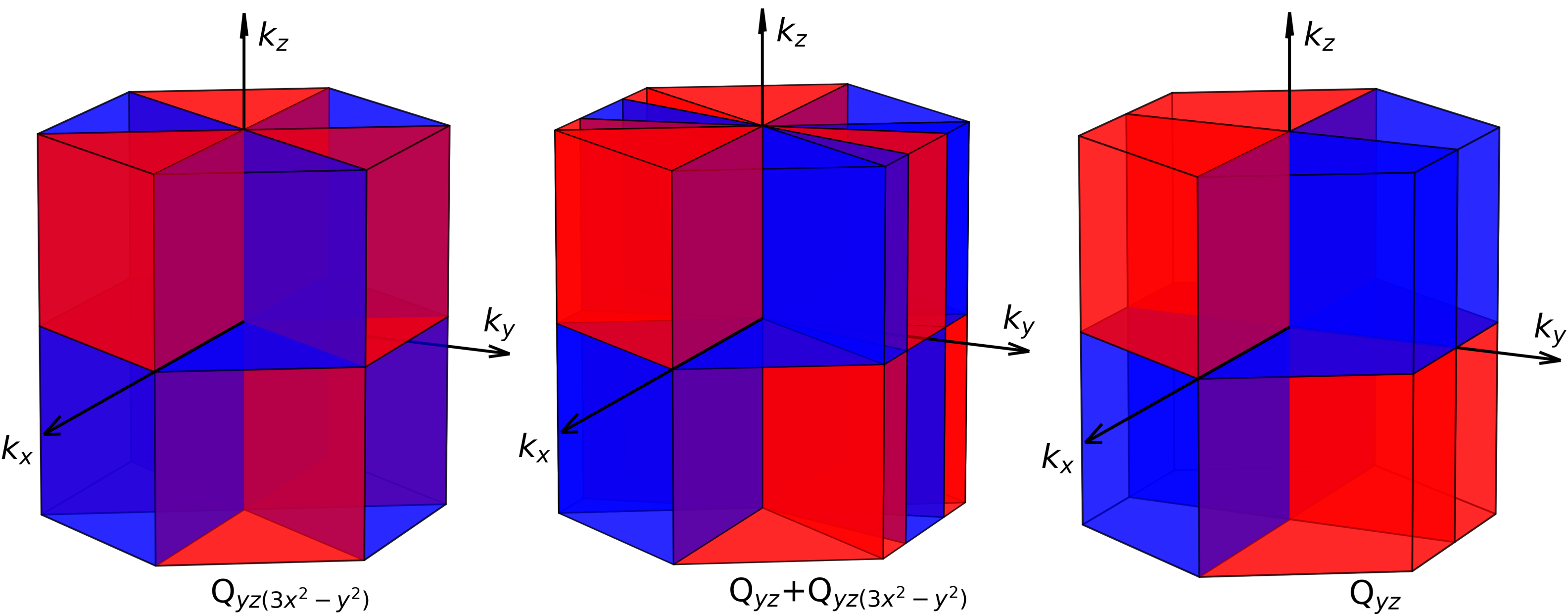}
    \caption{Crossover from the $g$-wave spin-momentum locking at $\lambda$=0 (left figure) to the $d$-wave (right figure).
    The top of the valence band of the centrosymmetric MnTe is represented by the middle figure, where both components are present.  }
    \label{fig:crossover}
\end{figure*}%S17

\section{Equations for the Rashba spin-momentum locking in the case of wurtzite M\lowercase{n}T\lowercase{e}}

The Rashba Hamiltonian $\mathcal{H}_{R}$ for an electric field applied along the $z$ axis can be written as
\begin{equation}\label{eq:Rashba}
\mathcal{H}_{R}=({\alpha_y}\sin{k_y}\,S_x - {\alpha_x}\sin{k_x}\,S_y)\,\sigma_0^{\mathrm{site}},
\end{equation}
where ${\alpha_y}$ and ${\alpha_x}$ are coefficients that determine the strength of the Rashba spin--orbit coupling. These parameters can be quantitatively determined by fitting the relativistic band structure~\cite{PhysRevB.109.115141}.  
The term proportional to $\sin{k_y}$ generates spin--momentum locking $Q_y$ for the associated $S_x$ spin component, while the term proportional to $\sin{k_x}$ generates spin--momentum locking $Q_x$ for the corresponding $S_y$ spin component. Therefore, the Hamiltonian in Eq.~\eqref{eq:Rashba} captures the Rashba contribution to the relativistic spin splitting in the case of wurtzite MnTe. Consequently, an analytical treatment of the RSML becomes possible using the methodology developed in the literature~\cite{Fakhredine25b}.

\section{Crossover from \lowercase{$g$}- to \lowercase{$d$}-wave in relativistic M\lowercase{n}T\lowercase{e} with N\lowercase{i}A\lowercase{s} structure}

In this section, we describe in detail how the non-relativistic $g$-wave altermagnet MnTe in the NiAs structure evolves into a $d$-wave state. We analyze the crossover from $g$-wave to $d$-wave behavior, focusing on the dominant spin component $S_y$. This crossover is illustrated graphically in Fig.~\ref{fig:crossover}.
In the left panel, we show the $g$-wave spin--momentum locking in the non-relativistic case with spin--orbit coupling $\lambda = 0$. The $g$-wave exhibits four nodal planes and can be described in momentum space by the magnetic hexapole $Q_{yz(3x^2 - y^2)}$. These four nodal planes are obtained by setting the argument of the multipole to zero, yielding the conditions $k_y = 0$, $k_z = 0$, and $3k_x^2 - k_y^2 = 0$.
When spin--orbit coupling is introduced ($\lambda \neq 0$) with the N\'eel vector aligned along the $y$-direction, the two nodal planes defined by $3k_x^2 - k_y^2 = 0$ are no longer symmetry-protected, but they cannot disappear instantly. Therefore, the nodal planes become accidental nodal surfaces where there is zero-spin splitting, but their position in the k-space is not symmetry-protected. Therefore, the multipole with the lowest wave allowed in the system is $Q_{yz}$; however, due to the reminiscence of the non-relativistic state, the $g$-wave remains relevant. In this regime, the spin--momentum locking of $S_y$ can be described as a linear combination of two contributions, $Q_{yz(3x^2 - y^2)}$ and $Q_{yz}$. This magnetic state is shown in the middle panel of Fig.~\ref{fig:crossover} and captures the properties at the top of the valence band of MnTe in the NiAs structure.
Upon further increasing the spin-orbit coupling or in bands where its effect is stronger, the $g$-wave component may be suppressed entirely. The two accidental nodal planes then merge at $k_x = 0$ and annihilate pairwise, leaving the system in a pure $d$-wave state described by the magnetic quadrupole $Q_{yz}$, as illustrated in the right panel of Fig.~\ref{fig:crossover}.

\renewcommand{\bibsection}

{\section*{References}}
\bibliographystyle{apsrev4-1}
\FloatBarrier
\bibliography{references}